\documentclass[nofootinbib,floatfix,prd,preprintnumbers,twocolumn,superscriptaddress]{revtex4}
\usepackage{graphicx}
\usepackage{amsfonts}
\usepackage{amssymb}
\usepackage{amsbsy}
\usepackage{amsmath}
\usepackage{latexsym}
\usepackage{bm}
\usepackage{hyperref}

\newcommand*{\di}{\partial}

\newcommand*{\rhohat}{\hat{\rho}}

\newcommand*{\z}{\hat{z}}
\renewcommand*{\c}{\text{c}}
\renewcommand*{\u}{\hat{u}}
\renewcommand*{\v}{\hat{v}}

\renewcommand*{\t}{\hat{t}}
\newcommand*{\y}{\hat{y}}
\renewcommand*{\a}{\hat{a}}
\renewcommand*{\b}{\text{b}}

\renewcommand*{\k}{\hat{k}}

\renewcommand*{\H}{\hat{H}}

\newcommand*{\approaches}[2]{\xrightarrow[#2]{\,\,\,{#1}\,\,\,}}

\begin{document}

\preprint{arXiv:0711.2563 [astro-ph]}

\title{Cosmological perturbations in the DGP braneworld: numeric solution}

\author{Antonio Cardoso}%
\email{antonio.cardoso.AT.port.ac.uk}%
\affiliation{Institute of Cosmology \& Gravitation, University of
Portsmouth, Portsmouth~PO1~2EG, UK}

\author{Kazuya Koyama}
\email{kazuya.koyama.AT.port.ac.uk}%
\affiliation{Institute of Cosmology \& Gravitation, University of
Portsmouth, Portsmouth~PO1~2EG, UK}

\author{Sanjeev S.~Seahra}
\email{sanjeev.seahra.AT.port.ac.uk} %
\affiliation{Institute of Cosmology \& Gravitation, University of
Portsmouth, Portsmouth~PO1~2EG, UK}%
\affiliation{Department of Mathematics \& Statistics, University of
New Brunswick \\ Fredericton, New Brunswick, Canada E3B 5A3}

\author{Fabio P.~Silva}
\email{fabio.silva.AT.port.ac.uk} %
\affiliation{Institute of Cosmology \& Gravitation, University of
Portsmouth, Portsmouth~PO1~2EG, UK}

\date{December 9, 2007}

\begin{abstract}

We solve for the behaviour of cosmological perturbations in the
Dvali-Gabadadze-Porrati (DGP) braneworld model using a new numerical
method.  Unlike some other approaches in the literature, our method
uses no approximations other than linear theory and is valid on
large scales.  We examine the behaviour of late-universe density
perturbations for both the self-accelerating and normal branches of
DGP cosmology.  Our numerical results can form the basis of a
detailed comparison between the DGP model and cosmological
observations.

\end{abstract}

\maketitle

\section{Introduction}

In the braneworld paradigm our universe is a lower-dimensional
object embedded in a higher-dimensional bulk spacetime. Standard
Model fields are assumed to be confined to the brane while gravity
is allowed to propagate in the bulk. The Dvali-Gabadadze-Porrati
(DGP) \cite{Dvali:2000rv,Dvali:2000hr} model postulates that we live
in a 4-dimensional hypersurface in a 5-dimensional Minkowski bulk.
General Relativity (GR) is recovered at small scales (smaller than
the crossover scale $r_\c$) due to the inclusion of an induced
gravity term in the action.

It was quickly realized that this model has two distinct classes of
cosmological solutions \cite{Deffayet:2000uy}. One of them exhibits
accelerated expansion at late times without the need to include any
exotic cosmological fluids, such as dark energy, or any brane
tension that acts as an effective 4-dimensional cosmological
constant. Hence, this branch of solutions is called
``self-accelerating''. Several attempts to confront the
self-accelerating universe with observations have been made
\cite{Fairbairn:2005ue, Maartens:2006yt, Song:2006jk} (also see
\cite[review]{Koyama:2007rx} and references therein). To explain the
observed acceleration we require $r_\c \sim H_0^{-1}$, where $H_0$
is the current value of the Hubble parameter.  It is expected that
structure formation will help to distinguish the self-accelerating
DGP universe from dark energy models based on 4-dimensional GR. This
is because the growth of cosmological perturbations is very
sensitive to the existence of an extra dimension.  A full
5-dimensional treatment is required to model these perturbations,
which is why obtaining observational predictions for the behaviour
of fluctuations in the DGP model is technically challenging.

Several authors have considered the problem of the dynamics of
perturbations in the DGP model, but they have all relied on some
sort of approximation or simplifying \emph{ansatz}.  Two examples of
this are the quasi-static (QS) approximation developed by
\citet{Koyama:2005kd} and the dynamical scaling (DS) \emph{ansatz}
proposed by \citet{Sawicki:2006jj}.  The former approximation scheme
solves the perturbative equations of motion by focussing on the
extreme subhorizon regime.  In contrast the DS method, which assumes
that perturbations evolve as power laws of the scale factor with
time-varying power law indices, is supposed to be valid on all
scales.  It has been shown that the DS solution approaches the QS
solution for subhorizon perturbations.


In this paper we present a complete numerical analysis of the
evolution of scalar perturbations in the DGP model.  Mathematically,
the problem involves the solution of a partial differential equation
in the bulk coupled to an ordinary differential equation on the
brane.  A numerical method for dealing with such systems has
previously been developed for cosmological perturbations in the
Randall-Sundrum (RS) model \cite{Cardoso:2006nh,Cardoso:2007zh}.
However, the DGP problem is more complicated than the RS case due to
a non-local boundary condition on the bulk field. Hence, the
algorithm used in this paper represents a significant generalization
of the one used in Refs.~\cite{Cardoso:2006nh,Cardoso:2007zh}.


Unfortunately, some theoretical issues cast doubt on the validity of
the self-accelerating DGP solutions.  Specifically, the existence of
a perturbative ghost perhaps suggests that this solution cannot
describe our Universe \cite{Koyama:2005tx, Gorbunov:2005zk}, though
there has been a debate on the physical implication of the ghost
mode (see \cite[review]{Koyama:2007za} and references therein).
However, as alluded to above there is another ``normal'' branch of
solutions in the DGP model. We cannot explain the late time
accelerated expansion of the Universe using the normal branch
without including an effective cosmological constant induced by the
brane tension $\sigma$. However, by allowing a non-zero $\sigma$,
normal branch solutions can mimic dark energy models with the
equation of state $w$ smaller than $-1$ \cite{Sahni:2002dx,
Lue:2004za}. Cosmological constraints on the background dynamics of
these models have been studied \cite{Lazkoz:2006gp, Lazkoz:2007zk}.
Unlike 4-dimensional models that realize $w < -1$ by the
introduction of a phantom field, the normal branch of DGP cosmology
is ghost-free \cite{Charmousis:2006pn}. This unique feature is what
motivates us to numerically study the perturbations of the normal
branch of DGP cosmology in the penultimate section of this paper.
During the preparation of this manuscript, we learned that the
behaviour of perturbations in the normal branch has been
independently obtained by \citet{Song0} using the DS method.


The structure of the paper is as follows: The background cosmology
of the DGP model is discussed in \S\ref{sec:background}.  In
\S\ref{sec:master}, we express the equations of motion for scalar
and tensor perturbations of the DGP model in the dimensionless
canonical form introduced in \S\ref{sec:canonical}. The numerical
method used to solve these canonical equations of motion is
developed in \S\ref{sec:numeric}.  Our algorithm is tested in
\S\ref{sec:tensor results}, where we numerically recover analytic
results for the behaviour of tensor perturbations in matter-free DGP
models.  In \S\ref{sec:scalar SA} and \S\ref{sec:scalar NB}, we
solve the scalar perturbations problem in the self-accelerating and
normal branches, respectively.  Finally, \S\ref{sec:conclusions} is
reserved for our conclusions.

\section{Background solution}\label{sec:background}

\subsection{Field equations and junction conditions}

We consider a 5-dimensional manifold $\mathcal{M}$ with metric
$g_{ab}$, and covered by coordinates $\{X^a\}_{a=0}^4$.  The
manifold has a 4-dimensional brane boundary $\di\mathcal{M}_\b$ with
intrinsic metric $\gamma_{\alpha\beta}$, and covered by coordinates
$\{x^\alpha \}_{\alpha=0}^3$. The action of the model is
\begin{multline}
    S =  \frac{1}{2\kappa_5^2} \int\limits_\mathcal{M} d^5 X \sqrt{-g} R^{(5)}
    + \frac{1}{2\kappa_4^2} \int\limits_{\di\mathcal{M}_\b} d^4 x \sqrt{-\gamma} R^{(4)} \\ +
    \int\limits_{\di\mathcal{M}_\b} d^4 x \sqrt{-\gamma} (\mathcal{L}_m
    - \sigma).
\end{multline}
Here, $\sigma$ is the brane tension and $\mathcal{L}_m$ is the
matter Lagrangian.  We impose the $\mathbb{Z}_2$ symmetry that the
bulk is mirror symmetric about the brane.\footnote{Technically,
$\mathcal{M}$ refers to one half of the total bulk spacetime.} The
field equations satisfied in $\mathcal{M}$ are simply
\begin{equation}
    R^{(5)}_{ab} = 0.
\end{equation}
We write the brane normal pointing \emph{into} $\mathcal{M}$ as
$n^a$. We find that the brane's extrinsic curvature,
\begin{equation}
    K_{\alpha\beta} = e^a_\alpha e^b_\beta \nabla_a n_b, \quad
    e^a_\alpha = \frac{\di X^a}{\di x^\alpha},
\end{equation}
must satisfy
\begin{equation}\label{eq:junction conditions}
    K_{\alpha\beta} - K \gamma_{\alpha\beta} - r_\c G_{\alpha\beta}^{(4)}
    = -\tfrac{1}{2} \kappa_5^2 (T_{\alpha\beta}-\sigma
    g_{\alpha\beta}),
\end{equation}
where we have defined the cross-over distance $r_\c$ by
\begin{equation}
    r_\c = \frac{\kappa_5^2}{2\kappa_4^2}.
\end{equation}
We assume that the stress-energy tensor of the brane matter,
\begin{equation}
    T_{\alpha\beta} = -\frac{2}{\sqrt{-\gamma}}
    \frac{\delta(\sqrt{-\gamma}\mathcal{L}_m)}{\delta
    \gamma^{\alpha\beta}} = (\rho + p)u_\alpha u_\beta + p
    g_{\alpha\beta},
\end{equation}
is of the perfect fluid form.  Note that as in GR, the stress energy
tensor is conserved, $\nabla^\alpha T_{\alpha\beta} = 0$.  Using the
Gauss-Codazzi equations, it is possible to re-write the junction
conditions (\ref{eq:junction conditions}) as the effective Einstein
equations
\begin{equation}
\label{eq:4D Einstein equations} G^{(4)}_{\mu\nu} =(2 \kappa_4^2
r_\c)^2 \Pi_{\mu \nu}-\mathcal{E}_{\mu\nu},
\end{equation}
where
\begin{gather}\nonumber
    \Pi_{\mu \nu} =  -\tfrac{1}{4} \tilde{T}_{\mu \alpha}
    \tilde{T}_{\nu}{}^{\alpha} +\tfrac{1}{12} \tilde{T}
    \tilde{T}_{\mu \nu} + \tfrac{1}{24}( 3 \tilde{T}_{\alpha \beta}
    \tilde{T}^{\alpha \beta}- \tilde{T}^2 )
    g_{\mu \nu}, \\
    \tilde{T}_{\mu \nu} =  T_{\mu \nu} - \sigma g_{\alpha\beta}
    - \kappa_4^{-2} G^{(4)}_{\mu \nu},
\end{gather}
and $\mathcal{E}_{\mu \nu}$ is the trace-free projection of the
5-dimensional Weyl tensor.

An important tool that is often used to analyze braneworld models is
the ``Gaussian-normal'' coordinates.  These are constructed by
looking at the spatial geodesics that extend perpendicularly from
$\di\mathcal{M}_\b$ into $\mathcal{M}$.  Gaussian-normal coordinates
are then given by $(x^\alpha,y)$, where $y$ is the affine parameter
along these geodesics and $x^\alpha$ are 4-dimensional coordinates
on the family of hypersurfaces tangent to the brane.  Without loss
of generality, we can set the brane to be at $y = 0$.  Most
importantly, the derivative of any bulk scalar quantity with respect
to $y$ on the brane corresponds to the normal derivative:
\begin{equation}
    (\di_y \psi)_\b = \frac{\di\psi}{\di y} \bigg|_{y=0} = n^a
    \di_a \psi.
\end{equation}
Below, $(\di_y \cdots)_\b$ will always be understood to be the
normal derivative of some quantity evaluated at the brane.

\subsection{Bulk geometry and brane trajectory}

One solution of the above field equations makes use of the following
5-dimensional flat metric with $R_{abcd} = 0$:
\begin{equation}\label{eq:bulk metric}
    ds^2 = g_{ab} dX^a dX^b = -r_\c^2\, du \, dv + v^2 \,d\mathbf{x}^2.
\end{equation}
Here, $u$ and $v$ are dimensionless null coordinates.  The brane is
defined parametrically by $(u,v) = (u_\b(t),v_\b(t))$, where
\begin{equation}
    v_\b(t) = a(t), \quad
    u_\b(t) = \frac{1}{r_\c^2} \int^t_0 \frac{dx}{\dot{a}(x)}.
\end{equation}
Here, $a(t)$ is the scale factor of the brane universe, normalized
to unity today, and $t$ is the proper time along the brane.  The
latter implies that the following induced line element on the brane
is of the FRW form:
\begin{equation}
    ds_\b^2 = \gamma_{\alpha\beta} dx^\alpha dx^\beta = -dt^2 + a^2(t)\,d\mathbf{x}^2.
\end{equation}
We have selected the $u$ coordinate such that $u_\b(0) = 0$ and the
brane moves in the direction of increasing $u$ and $v$.  The
5-velocity of the brane $u^a$ and brane normal $n^a$ are given by
\begin{subequations}
\begin{align}
    \di_t & = u^a \di_a = \frac{dx^a}{dt} \frac{\di}{\di x^a} = \frac{1}{r_\c^2
    \dot{a}}\di_u + \dot{a}
    \, \di_v, \\ \di_y & = n^a \di_a = \epsilon \left[ -\frac{1}{r_\c^2 \dot{a}} \di_u + \dot{a}
    \di_v \right] \label{eq:normal def}.
\end{align}
\end{subequations}
These satisfy
\begin{equation}
    u^a u_a = -1, \quad u^a n_a = 0, \quad n^a n_a = 1.
\end{equation}
The $\epsilon = \pm 1$ parameter in the definition of the brane
normal (\ref{eq:normal def}) reflects the fact that when we impose
the $\mathcal{Z}_2$ symmetry across the brane, we have two choices
for the half of the bulk manifold $\mathcal{M}$ we discard.

The junction conditions (\ref{eq:junction conditions}) can be used
to determine the brane dynamics, which are governed by
\begin{subequations}\label{eq:background dynamics}
\begin{align}
    H & = \frac{\dot{a}}{a} = \frac{1}{2r_\c} \left[ \epsilon + \sqrt{1 + \frac{4}{3}\kappa_4^2 r_\c^2
    (\rho+\sigma)}\right], \\
    \frac{d\rho}{dt} & = -3(1+w)\rho H, \\
    \frac{dH}{dt} & = - \frac{r_\c \kappa_4^2 (1+w)
    \rho H}{\sqrt{1+\frac{4}{3} r_\c^2 \kappa_4^2(\rho+\sigma)}}.
\end{align}
\end{subequations}
Note that when $\epsilon = +1$, we have that $Hr_\c \approx 1$
when the density of brane matter is small $|\rho + \sigma| \ll
\kappa_4^{-2} r_\c^{-2}$.  This implies a late-time accelerating
universe, which is why the $\epsilon = +1$ case is called the
self-accelerating branch and the $\epsilon = -1$ case is called
the normal branch.

\section{Master equations governing perturbations}\label{sec:master}

\subsection{Dimensionless coordinates and canonical wave
equations}\label{sec:canonical}

In this paper, we will consider the perturbations of the DGP model
governed by a field $\psi$ defined on the bulk spacetime
$\mathcal{M}$ coupled to a dynamical field $\Delta$ residing on the
brane $\di\mathcal{M}_\b$.  It is useful to decompose these fields
into Fourier modes as follows:
\begin{subequations}
\begin{align}
    \psi(u,v,\mathbf{x}) & = \int \frac{d^3 k}{(2\pi)^{3/2}}
    \psi_\mathbf{k}(u,v) e^{i\mathbf{k}\cdot\mathbf{x}}, \\
    \Delta(t,\mathbf{x}) & = \int \frac{d^3 k}{(2\pi)^{3/2}}
    \Delta_\mathbf{k}(t) e^{i\mathbf{k}\cdot\mathbf{x}}.
\end{align}
\end{subequations}
As usual for linear theory, the individual $\mathbf{k}$ modes are
decoupled from one another.  In what follows, we will omit the
$\mathbf{k}$ subscript from $\psi_\mathbf{k}$ and
$\Delta_\mathbf{k}$.  That is, $\psi$ and $\Delta$ refer to the
Fourier amplitudes of modes with wavevector $\mathbf{k}$.  We will
also assume that a normalization has been selected such that the
Fourier amplitudes are dimensionless.

For a given mode, we define the ``*'' epoch as the moment when a
given mode crosses the Hubble horizon:
\begin{equation}
    k = H_* a_*, \quad k^2 = \mathbf{k} \cdot \mathbf{k}.
\end{equation}
Then, we can define a set of normalized variables (decorated with
hats):
\begin{align}\nonumber
    \H & = Hr_\c, &
    \k & = k r_\c/a_*, &
    \rhohat & = \kappa_4^2 r_\c^2 \rho, \\ \nonumber
    \hat\sigma & = \kappa_4^2 r_\c^2 \sigma, &
    \t & = {t}/{r_\c}, &
    \y & = {y}/{r_\c}, \\
    \a & = {a}/{a_*}, &
    \u & = a_* u, &
    \v & = {v}/{a_*}. \label{eq:dimensionless quantities}
\end{align}
Explicitly, we have
\begin{subequations}\label{eq:H and k def}
\begin{align}
    \H & = \frac{1}{2}
    \left[ \epsilon + \sqrt{1 + \frac{4}{3}(\rhohat + \hat{\sigma}) } \right],
    \\ \k & = \frac{1}{2} \left[ \epsilon + \sqrt{ 1 +
    \frac{4}{3}(\rhohat_*+\hat\sigma)
    } \right] = \H_*.
\end{align}
\end{subequations}
It is also useful to define the 2-dimensional Cartesian
coordinates
\begin{equation}
    x^0 = \tau = \frac{\v + \u}{2}, \quad x^1 = z =
    \frac{\v-\u}{2}.
\end{equation}
In terms of these coordinates, the dimensionless tangential
$\hat\di_t = U^A \di_A$ and normal $\hat\di_y = N^A \di_A$
derivatives to the brane are
\begin{subequations}
\begin{align}
    \hat\di_t = \frac{1}{2} \left[
    \left( \H\a + \frac{1}{\H\a} \right) \frac{\di}{\di\tau}
    + \left( \H\a - \frac{1}{\H\a} \right) \frac{\di}{\di z}
    \right], \\
    \hat\di_y = \frac{\epsilon }{2} \left[
    \left( \H\a - \frac{1}{\H\a} \right) \frac{\di}{\di\tau}
    + \left( \H\a + \frac{1}{\H\a} \right) \frac{\di}{\di z}
    \right],
\end{align}
\end{subequations}
where
\begin{equation}
    -1 =
    U_A U^A, \quad 1 =
    N_A N^A, \quad 0 =
    U_A N^A,
\end{equation}
and capital roman indices $A,B=0,1$ are raised and lowered with
the flat 2-metric $\eta_{AB} = \text{diag}(-1,1)$.

The brane trajectory in the $(\tau,z)$ plane is given by the
solution of
\begin{subequations}\label{eq:brane EOM}
\begin{align}
    \frac{d\a}{d\t} & = \H\a, \\
    \frac{d\tau_\b}{d\t} & = \frac{1}{2}
    \left(  \H\a + \frac{1}{\H\a} \right), \\
    \frac{dz_\b}{d\t} &= \frac{1}{2} \left(  \H\a - \frac{1}{\H\a} \right),
\end{align}
\end{subequations}
subject to the initial conditions
\begin{equation}
    \a(0) =
    \a_i, \quad \tau_\b(0) = \tfrac{1}{2} \a_i, \quad z_\b(0) =
    \tfrac{1}{2} \a_i.
\end{equation}
If $\epsilon = +1$, the normal points in the direction of increasing
$z$ and the ``bulk'' corresponds to the the portion of the
$(\tau,z)$ plane to the right of the brane; if $\epsilon = -1$ the
opposite is true.

We will find below that the equations governing the bulk $\psi =
\psi(\tau,z)$ and brane field $\Delta = \Delta(\t)$ are of the form:
\begin{subequations}\label{eq:dimensionless system}
\begin{align}
    \label{eq:psi EOM} 0 & =
    (\di_\tau^2 - \di_z^2 + V) \psi,\\
    \label{eq:boundary condition}(\di_{\hat{y}} \psi)_\b & =  \lambda_1 \Delta + \lambda_2
    \Xi
    + \lambda_3 \psi_\b + \lambda_4\psi_\b' + \lambda_5 \psi_\b'', \\
    \Xi' & = \lambda_6 \Delta + \lambda_7 \Xi + \lambda_8 \psi_\b +
    \lambda_9 \psi_\b' + \lambda_{10} \psi_\b'', \label{eq:Delta EOM} \\
    \Delta' & = \Xi, \label{eq:Xi def}
\end{align}
\end{subequations}
where we have introduced the auxiliary field $\Xi$, which
corresponds to the time derivative of $\Delta$.  In
(\ref{eq:dimensionless system}), a prime $'$ denotes the derivative
with respect to the dimensionless proper time $d/d\t$ and $\psi_\b =
\psi_\b(\t) = \psi(\tau_\b(\t),z_\b(\t))$ is the value of the bulk
field on the brane.  Also, $V = V(\tau,\z)$ is the bulk potential
and the coefficients $\lambda_n = \lambda_n(\a)$ are functions of
the brane scale factor.  All quantities appearing in
(\ref{eq:dimensionless system}) are dimensionless.  We refer to
(\ref{eq:dimensionless system}) as the canonical form of the
perturbative equations of motion.

\subsection{Density perturbations in the late universe}

In this subsection we describe the formulae governing scalar
perturbations in the late-time matter dominated universe.  We take
the matter content of the brane to be a dust fluid $w = 0$ (i.e.,
cold dark matter):
\begin{equation}
    \rho \propto a^{-3}.
\end{equation}

\begin{table*}
\begin{ruledtabular}
\begin{tabular}{cll}

coefficient &

\multicolumn{1}{c}{scalar case (matter domination)} &

\multicolumn{1}{c}{tensor case (pure tension brane)}  \\ \hline
\\[-8pt]

$\lambda_1$ & $\tfrac{3}{2}\epsilon \rhohat \a^{3/2} \k^{-2} \gamma_4 $ & $0$ \\

$\lambda_2$ & $0$ & $0$ \\

$\lambda_3$ & $-\tfrac{3}{4}\epsilon \H^{-1} \H' \gamma_1 - \frac{3}{8}\epsilon \H
(4 + 3\gamma_1 - 9\gamma_3 +2\epsilon \gamma_4) - \frac{3}{4} \epsilon \k^2 \H^{-1}
\a^{-2} \gamma_3 $ & $\k^2 \a^{-2} + \frac{3}{4}\H(2\epsilon-\H)$ \\

$\lambda_4$ & $\frac{3}{4} \epsilon (3\gamma_3 - 2\gamma_2)$ & $0$ \\

$\lambda_5$ & $-\frac{1}{2} \epsilon \H^{-1} \gamma_1$ & $1$ \\

$\lambda_6$ & $\tfrac{1}{2} \rhohat \gamma_2$ & $0$ \\

$\lambda_7$ & $-2\H$ & $0$ \\

$\lambda_8$ & $-\tfrac{1}{4}\epsilon\k^4 \a^{-7/2}\gamma_4$ & $0$ \\

$\lambda_9$ & $0$ & $0$ \\

$\lambda_{10}$ & $0$ & $0$ \\

$\lambda_{11} = \lambda'_5$ & $\frac{1}{2} \epsilon \H^{-2} \H' \gamma_1$ & $0$ \\

$\lambda_{12} = \lambda'_{10}$ & $0$ & $0$ \\

\end{tabular}
\end{ruledtabular}

\caption{Dimensionless coefficients appearing in the canonical wave
equations (\ref{eq:dimensionless system}) for the case of scalar
perturbations in the late universe and tensor perturbation of a pure
tension brane.  Even though the $\lambda_{11}$ and $\lambda_{12}$
coefficients do not appear in (\ref{eq:dimensionless system}), they
are crucial for the numeric scheme developed in \S\ref{sec:numeric}.
We make use of the $\gamma$-factors defined in Eq.~(\ref{eq:gamma
defs}) and the following notation: $\H' = d\H/d\t = - (1+w) \rhohat
(1+\frac{4}{3} \hat{\mu})^{-1/2}$, $\hat\mu = \rhohat + \hat\sigma$,
and $ \rhohat = \rhohat_* \a^{-3}$.}\label{tab:coefficients}

\end{table*}
In the 5-dimensional longitudinal gauge, scalar-type
perturbations\footnote{That is, perturbations that can be derived
from scalar potentials defined on the 3-dimension spatial
sections.} of the bulk geometry (\ref{eq:bulk metric}) can be
written as
\begin{multline}
    ds^2 = -r_\c^2 ( du \, dv + F_{uu} du^2 + 2F_{uv} du \,dv +
    F_{vv} dv^2 ) \\ + r_\c (f_{ui} du + f_{vi} dv ) dx^i + v^2
    (1 + 2\mathcal{R}) \delta_{ij} dx^i dx^j.
\end{multline}
It can be shown that the dynamics of all of the perturbative
quantities in this expression can be derived from a single scalar
bulk degree of freedom \cite{Mukohyama:2000ui}.  After Fourier
decomposition, we find that the mode amplitude $\Omega =
\Omega(u,v)$ of this master field obeys
\begin{equation}\label{eq:scalar master eqn}
    0 = \frac{\di^2 \Omega}{\di u \di v} - \frac{3}{2v} \frac{\di\Omega}{\di
    u} + \frac{k^2 r_\c^2}{4v^2} \Omega.
\end{equation}
We parameterize the fluctuations of the brane geometry by the two
metric potentials $\Phi$ and $\Psi$:
\begin{equation}\label{eq:perturbed induced metric}
    ds_\b^2 = -(1+2\Psi)dt^2 + a^2 (1+2\Phi)\delta_{ij} dx^i
    dx^j.
\end{equation}
Finally, we write the perturbation of the brane fluid stress
energy tensor as
\begin{equation}\label{eq:perturbed stress energy tensor}
    \delta T^0{}_0 = -\delta\rho, \quad \delta T^{0}{}_i = a \,
    \di_i \delta q, \quad \delta T^i{}_j = \delta p \, \delta^i{}_j.
\end{equation}
One can construct the comoving density perturbation from these
quantities as follows:
\begin{equation}\label{eq:Delta defn}
    \rho \Delta =
    \delta\rho - 3Ha \, \delta q.
\end{equation}
Both $\Delta$ and $\Omega$ are gauge invariant quantities under a
coordinate transformation on the brane and in the bulk.  Note that
$\Delta$ is the density contrast of the cold dark matter only, not
the the total density contrast.

As shown in \cite{Deffayet:2002fn} and demonstrated in Appendix
\ref{sec:scalar derivation}, one can use the effective Einstein
equations on the brane and the detailed relationship between
$\Omega$ and the metric perturbations to find the following boundary
condition for $\Omega$:
\begin{multline}\label{eq:scalar boundary condition}
    (\di_y \Omega)_\b = - \frac{\epsilon \gamma_1}{2H}
    \ddot\Omega_\b +\frac{9\epsilon \gamma_3}{4} \dot\Omega_\b -
    \\
    \frac{3(\epsilon\gamma_3 k^2 + \gamma_4 H^2 a^2) }{4H a^2}
    \Omega_\b + \frac{3\epsilon r_\c\kappa_4^2 \rho a^3 \gamma_4 }{2k^2}
    \Delta;
\end{multline}
the following equation of motion for $\Delta$:
\begin{equation}\label{eq:scalar brane equation}
    \ddot \Delta + 2H \dot \Delta - \frac{1}{2} \kappa_4^2 \rho
    \gamma_2 \Delta = - \frac{\epsilon \gamma_4 k^4}{4a^5}
    \Omega_\b;
\end{equation}
and the following expressions for $\Phi$ and $\Psi$:
\begin{eqnarray}\nonumber
    \Phi & = & +\frac{\kappa_4^2 \rho a^2 \gamma_1}{2k^2}\Delta +
    \frac{\epsilon \gamma_1}{4ar_\c} \dot\Omega_\b -
    \frac{\epsilon(k^2+3H^2a^2)\gamma_1}{12 Hr_\c a^3} \Omega_\b,  \\
    \nonumber
    \Psi & = & -\frac{\kappa_4^2 \rho a^2 \gamma_2}{2k^2}\Delta +
    \frac{\epsilon \gamma_1}{4Hr_\c a } \ddot \Omega_\b -
    \frac{3\epsilon H \gamma_4}{4a} \dot\Omega_\b + \\ & &
    \frac{\epsilon(k^2 r_\c \gamma_4 + Ha^2 \gamma_2)}{4r_\c
    a^3}\Omega_\b, \label{eq:Phi and Psi formulae}
\end{eqnarray}
where $\Omega_\b = \Omega_\b(t) = \Omega(u_\b(t),v_\b(t))$. In
these expressions, the dimensionless $\gamma$-factors are:
\begin{subequations}\label{eq:gamma defs}
\begin{align}
    \gamma_1 & = \frac{2\epsilon H r_\c}{2\epsilon H r_\c - 1}, \\
    \gamma_2 & = \frac{2\epsilon r_\c(\dot H - H^2 + 2\epsilon H^3 r_\c)}
    {H(2\epsilon H r_\c -1)^2}, \\
    \gamma_3 & = \frac{4\epsilon r_\c(2\epsilon r_\c \dot H - 3H + 6\epsilon H^2 r_\c)}
    {9 (2\epsilon H r_\c -1)^2}, \\
    \gamma_4 & = \frac{4\epsilon (\epsilon r_\c \dot H - H + 2\epsilon H^2 r_\c)}
    {3 H (2\epsilon H r_\c -1)^2}.
\end{align}
\end{subequations}
From these formulae, it follows that the bulk field $\Omega$ has
dimensions of $(\text{length})^2$.  The bulk wave equation
(\ref{eq:scalar master eqn}), boundary condition (\ref{eq:scalar
boundary condition}) and (\ref{eq:scalar brane equation}) are the
equations we must solve.  Once we know $\Delta$ and $\Omega$ the
metric perturbations $\Phi$ and $\Psi$ can be obtained by
differentiation.  Another quantity of interest is the curvature
perturbation in uniform density slices, which is given by
\begin{equation}
    \zeta = \Phi + \frac{Ha}{\rho} \delta q + \frac{1}{3} \Delta,
\end{equation}
assuming matter domination.  This can be explicitly represented in
terms of $\Delta$ and $\Omega_\b$:
\begin{equation}\label{eq:zeta}
    \zeta = \left( \frac{1}{3}
    + \frac{\kappa_4^2 \rho a^2 \gamma_1}{2k^2}
    \right)
    \Delta + \frac{Ha^2}{k^2} \dot\Delta
    -\frac{\epsilon \gamma_1 k^2}{12 Hr_\c a^3} \Omega_\b.
\end{equation}
This quantity is interesting because it is expected to be conserved
on superhorizon scales for any metric theory of gravity
\cite{Wands:2000dp}, including the DGP model.  Hence, an explicit
verification that $\zeta$ is constant when $k \ll Ha$ provides a
useful consistency check of our numerical code.

We can define a dimensionless canonical bulk field $\psi$ by
\begin{equation}
    \psi = \frac{a_*^{1/2}}{v^{3/2} r_\c^2} \Omega.
\end{equation}
Substitution of this into (\ref{eq:scalar master eqn}) confirms
that $\psi$ satisfies the canonical bulk wave equation
(\ref{eq:psi EOM}) with potential
\begin{equation}\label{eq:scalar potential}
    V(\tau,z) = \frac{\k^2}{(\tau+z)^2}.
\end{equation}
We can then then replace $\Omega$ with $\psi$ in (\ref{eq:scalar
boundary condition}) and (\ref{eq:scalar brane equation}) and make
use of (\ref{eq:dimensionless quantities}) to obtain the canonical
boundary condition (\ref{eq:boundary condition}) and $\Delta$
equation of motion (\ref{eq:Delta EOM}). The explicit expressions
for the dimensionless $\lambda_n$ coefficients are given in Table
\ref{tab:coefficients}. Finally, it is useful to define a
dimensionless version of $\Omega$ as follows:
\begin{equation}\label{eq:Omega hat def}
    \hat\Omega = \hat{v}^{3/2} \psi = a_*^{-1} r_\c^{-2} \Omega.
\end{equation}
Generally speaking, it is more useful to work with $\hat\Omega$ than
$\Omega$ for numeric computations, and later on we will present
plots of $\hat\Omega$ instead of $\Omega$.

\subsection{Tensor perturbations}\label{sec:tensor mode eqns}

For tensor perturbations, we restrict ourselves to cases where there
is no ordinary matter on the brane,
\begin{equation}
    \rhohat = 0, \quad \hat\sigma = 3Hr_\c(Hr_\c - \epsilon).
\end{equation}
Tensor type perturbations of the bulk geometry are described by
perturbations of the form:
\begin{equation}
    ds^2 = -r_\c^2 \,du \, dv + v^2 (\delta_{ij} + E_{ij}) dx^i
    dx^j,
\end{equation}
where the 3-tensor $E_{ij}$ is given by
\begin{equation}
    E_{ij} = \sum_{A=\oplus,\otimes} \int \frac{d^3 k}{(2\pi)^{3/2}} E^A_\mathbf{k}(u,v)
    e^{i\mathbf{k}\cdot\mathbf{x}} e^A_{ij}(\mathbf{k}).
\end{equation}
Here, $e^A_{ij}$ is a constant polarization tensor satisfying
\begin{equation}
    \di_a e^A_{ij}(\mathbf{k}) =
    \delta^{ij} e^A_{ij}(\mathbf{k}) = k^i e^A_{ij}(\mathbf{k}) =
    0.
\end{equation}
For brevity, we omit the $\mathbf{k}$ subscript and $A$
superscript on the mode amplitude $E^A_\mathbf{k}(u,v)$ below.
Perturbations of the bulk Einstein equations yield
\begin{equation}
    \delta R_{ab}^{(5)} = 0 \quad \Rightarrow \quad
    0 = \frac{\di^2 E}{\di u \, \di v} + \frac{3}{2v} \frac{\di E}{\di
    u} + \frac{k^2 r_\c^2}{4v^2} E.
\end{equation}
Note that this is equivalent to $\nabla^a \nabla_a
(Ee^{i\mathbf{k}\cdot\mathbf{x}}) = 0$.  Perturbation of the
junction conditions yields that
\begin{equation}\label{eq:tensor boundary condition}
    (\di_y E)_\b = r_\c \left( \ddot E_\b + 3H \dot E_\b +
    \frac{k^2}{a^2} E_\b \right).
\end{equation}
To bring this into the canonical form we introduce the new bulk
variable
\begin{equation}\label{eq:tensor psi def}
    \psi = \frac{v^{3/2}}{a_*^{3/2}} E,
\end{equation}
which satisfies the canonical bulk wave equation with potential
\begin{equation}\label{eq:tensor potential}
    V(\tau,z) = \frac{\k^2}{(\tau+z)^2}.
\end{equation}
Substitution of (\ref{eq:tensor psi def}) into the boundary
condition (\ref{eq:tensor boundary condition}) yields the other
canonical coefficients $\lambda_n$ in (\ref{eq:dimensionless
system}), which are listed in Table \ref{tab:coefficients}.  Note
that since there is no brane field $\Delta$ in this case, we set
$\lambda_n = 0$ for $n = 1,2,6,7,8,9,10$.

\section{Numeric method}\label{sec:numeric}

\subsection{Computational grid}

\begin{figure*}
    \includegraphics[width=0.9\textwidth]{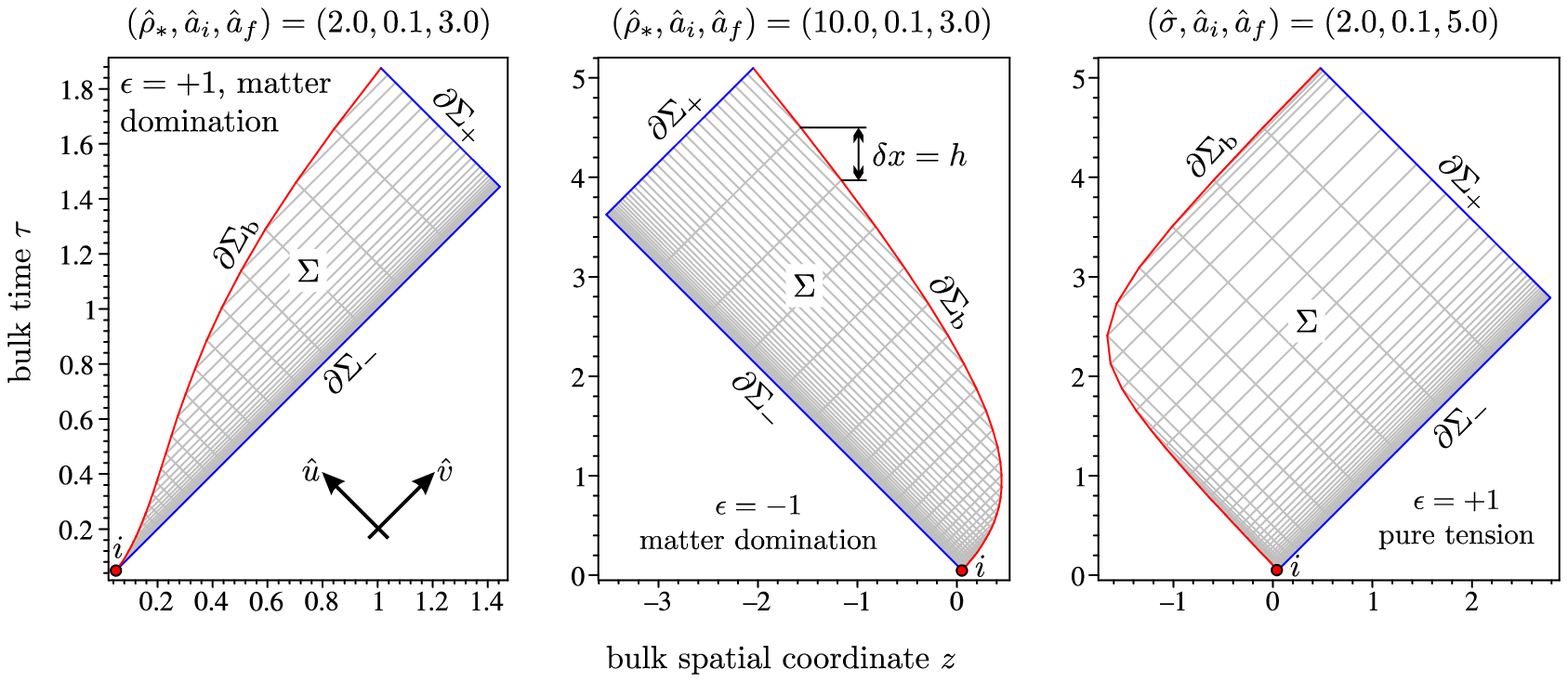}
    \caption{Typical computational grids used to solve the perturbation
    equations.
    }\label{fig:grids}
\end{figure*}

In this section, we develop a numeric algorithm to solve the system
of equations (\ref{eq:dimensionless system}) over a finite region
$\Sigma$ of the $(\tau,z)$ plane.  Illustrations of the shape of
$\Sigma$ are given in Fig.~\ref{fig:grids} for a few different brane
trajectories and choices of $\epsilon$.  The computational domain
has three distinct boundaries: the brane $\di\Sigma_\b$, a past null
hypersurfaces $\di\Sigma_-$, and a future null hypersurface
$\di\Sigma_+$.  Note how the computational domain is to the right of
$\di\Sigma_\b$ for the self-accelerating case $\epsilon = +1$ and to
the left for the normal case $\epsilon = -1$.  The brane size at the
beginning of the simulation (where $\di\Sigma_\b$ and $\di\Sigma_-$
intersect, marked $i$ in the figure) is $\a_i$, while the size at
the end (where $\di\Sigma_\b$ and $\di\Sigma_+$ intersect) is
$\a_f$.

Our numerical algorithm employs an irregular computational grid as
shown in Figs.~\ref{fig:grids} and \ref{fig:boundary cells}.  To
define this grid, we introduce an arbitrary time parametrization
along the brane given by the monotonically increasing function $x =
x(\t)$. The portion of the brane between $\a = \a_i$ and $\a = \a_f$
is subdivided into piecewise linear segments equally spaced in the
new $x$ parameter. That is, the change in $x$ over a given segment
is $\delta x = h$, where $h$ is the overall stepsize parameter of
the algorithm.  The bulk grid is then completed by drawing null
lines emanating from the endpoints of these segments as shown in
Fig.~\ref{fig:grids}.  As in previous work, the grid involves a
number of triangular cells adjacent to the brane and diamond shaped
cells in the bulk. The bulk cells will generally not be uniform in
size due to our choice of brane partitioning.

Our actual choice of $x = x(\t)$ is motivated by the desire to
obtain a quickly-converging algorithm that samples the expansion
history of the brane sufficiently densely.  We have experimented
with a number of possibilities and have found that $x(\t) = \ln
\tau_\b(\t)$ works best for the self-accelerating branch.  On the
other hand, $x(\t) = \tau_\b(\t)$ seems to give good performance for
the normal branch.  Of course, the choice of $x$ ultimately does not
matter since all possibilities should give the same results in the
$h \rightarrow 0$ limit.

\subsection{Evolution near the brane}\label{sec:triangle evolution}

\begin{figure}
    \includegraphics[width=0.95\columnwidth]{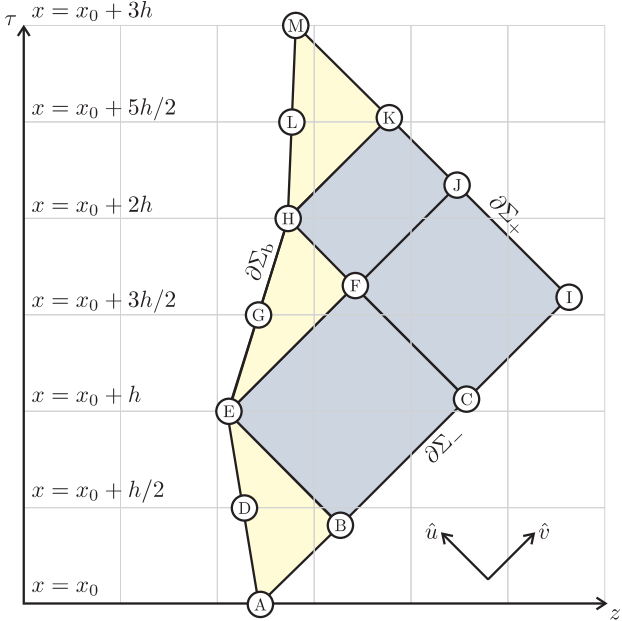}
    \caption{Grid geometry used to derive evolution formulae in
    \S\ref{sec:triangle evolution} and \S\ref{sec:diamond evolution}.
    The principal brane nodes A, E, H and M are separated by a brane
    time interval $\delta x = h$.  We have also introduced half-step
    nodes D, G and L, which are separated from the adjacent principal
    nodes by $\delta x = h/2$.  The half-step nodes are needed because
    of the non-local nature of the boundary condition in the DGP model.}
    \label{fig:boundary cells}
\end{figure}
In order to model the evolution of $\Delta$ and $\psi$ near the
brane, consider the geometry shown in Fig.~\ref{fig:boundary cells}.
Roughly speaking, our goal here is to develop an algorithm to
calculate the values of the fields at the nodes G and H given the
knowledge of their values at (A,D,E,F).

If we integrate the bulk wave equation over the triangular cell
EHF, we obtain
\begin{multline}\label{eq:triangle evolve 1}
    2 \psi_\text{F} -
    \psi_\text{H} - \psi_\text{E} = \int_\triangle d^2 x \,V\psi + \int_\text{E}^\text{H}
    d\t \, \di_{\hat{y}} \psi =
    \int_\triangle d^2 x \,V\psi + \\ \int_\text{E}^\text{H} d\t \, (\lambda_1 \Delta +
    \lambda_2 \Xi + \lambda_3 \psi_\b + \lambda_4 \psi_\b' + \lambda_5 \psi_\b'').
\end{multline}
Here, we have used (\ref{eq:boundary condition}) to substitute for
the normal derivative of $\psi$.  We can also integrate
(\ref{eq:Delta EOM}) and (\ref{eq:Xi def}) over the brane segment
from E to H, we get
\begin{gather}
    \nonumber \Xi_\text{H} - \Xi_\text{E} = \int_\text{E}^\text{H} d\t\,
    (\lambda_6 \Delta + \lambda_7 \Xi + \lambda_8 \psi_\b + \lambda_9 \psi_\b'+
    \lambda_{10} \psi_\b''), \\ \Delta_\text{H} - \Delta_\text{E}
     = \int_\text{E}^\text{H} d\t\, \Xi. \label{eq:triangle evolve 2}
\end{gather}

We now replace the integrals in the exact expressions
(\ref{eq:triangle evolve 1}) and (\ref{eq:triangle evolve 2}) with
discrete approximations. First, let us consider the 2-dimensional
integral in (\ref{eq:triangle evolve 1}).  A simple linear
approximation to $\psi$ inside EHF yields
\begin{equation}
    \int_{\triangle} d^2 x \, V \psi = \frac{\delta \t^2}{12} (V_\text{E} \psi_\text{E}
    + V_\text{H} \psi_\text{H} + V_\text{F} \psi_\text{F}) + \mathcal{O}(\delta \t^4).
\end{equation}
Here, $\delta \t$ is the proper time interval between the nodes E
and H, which is explicitly
\begin{equation}\label{eq:delta t integral}
    \delta \t = \int_\text{E}^\text{H} d\t = \int_\text{E}^\text{H}
    dx \, \Upsilon = \tfrac{1}{2} h \langle \Upsilon
    \rangle + \mathcal{O}(h^3).
\end{equation}
Here, we have introduced the notation
\begin{equation}
    \langle X \rangle = X_\text{H}
    + X_\text{E} , \quad \| X \| =
    X_\text{H} - X_\text{E},
\end{equation}
and defined
\begin{equation}
    \Upsilon = \frac{d\t}{dx} = \frac{d\t}{d\tau_\b}
    \frac{d\tau_b}{dx} = \left[ \frac{1}{2}
    \left(  \H\a + \frac{1}{\H\a} \right) \right]^{-1}
    \frac{d\tau_b}{dx}.
\end{equation}
The last equality in (\ref{eq:delta t integral}) follows from the
trapezoidal approximation for one-dimensional integrals.  We use
this same approximation for the other line integrals in
(\ref{eq:triangle evolve 1}) and (\ref{eq:triangle evolve 2}), after
integration by parts to remove the $\psi_\b''$ terms and a change of
variables from $\t$ to $x$.
It is useful to simplify the notation by introducing some new
coefficients:
\begin{align}\nonumber
    c_1    & = \tfrac{1}{2}\Upsilon \lambda_1, &
    c_6    & = \tfrac{1}{2}\Upsilon \lambda_6, &
    c_{11} & = \tfrac{1}{2} \Upsilon. \\ \nonumber
    c_2    & = \tfrac{1}{2}\Upsilon \lambda_2, &
    c_7    & = \tfrac{1}{2}\Upsilon \lambda_7, \\ \nonumber
    c_3    & = \tfrac{1}{2} \Upsilon \lambda_3, &
    c_8    & = \tfrac{1}{2} \Upsilon \lambda_8, \\ \nonumber
    c_4    & = \tfrac{1}{2}(\lambda_4 - \lambda'_5), &
    c_9    & = \tfrac{1}{2}(\lambda_9 - \lambda'_{10}), \\
    c_5    & = \Upsilon^{-1} \lambda_5, &
    c_{10} & = \Upsilon^{-1} \lambda_{10},
\end{align}
It is worthwhile noting that all of these $c_i$ coefficients are
functions of the brane trajectory only, and we assume that they are
known \emph{exactly}. In terms of these we get
\begin{subequations}\label{eq:triangle evolve 4}
\begin{eqnarray}
    \nonumber 2 \psi_\text{F} -
    \langle \psi \rangle & = &
    h \langle c_1 \Delta +
    c_2 \Xi + c_3 \psi_\b +
    c_4 \psi_\b^\bullet
    \rangle   +
    \| c_5 \psi_\b^\bullet \| + \\ & & \tfrac{1}{12} h^2 \langle
    c_{11} \rangle^2 [ \langle V\psi \rangle
    + V_\text{F} \psi_\text{F} ] + \mathcal{O}(h^3),
    \\ \nonumber \| \Xi \| & = &
    h \langle c_6 \Delta + c_7 \Xi + c_8 \psi_\b +
    c_9 \psi_\b^\bullet
    \rangle  + \|
    c_{10} \psi_\b^\bullet \| \\ & &  + \mathcal{O}(h^3), \\ \| \Delta \|
    & = & h \langle c_{11} \Xi \rangle +
    \mathcal{O}(h^3),
\end{eqnarray}
\end{subequations}
where
\begin{equation}
    \psi_\b^\bullet = \frac{d\psi_\b}{dx} = \frac{d\psi_\b}{d\t} \frac{d\t}{dx} =  \Upsilon \psi_\b'.
\end{equation}
We can now make use of the following approximations:
\begin{subequations}\label{eq:triangle evolve 5}
\begin{align}
    \psi_\text{E}^\bullet & = \frac{12\psi_\text{E}-16\psi_\text{D} + 3\psi_\text{A} +
    \psi_\text{H}}{6h} + \mathcal{O}(h^3), \\ \psi_\text{H}^\bullet & =
    \frac{-36\psi_\text{E}+32\psi_\text{D} -9 \psi_\text{A} +
    13\psi_\text{H}}{6h} + \mathcal{O}(h^3),
\end{align}
\end{subequations}
to eliminate the $\psi$ derivatives in (\ref{eq:triangle evolve 4})
to order $h^3$.  Once (\ref{eq:triangle evolve 5}) is substituted
into (\ref{eq:triangle evolve 4}), we have a linear system for
$(\psi_\text{H},\Delta_\text{H},\Xi_\text{H})$ in terms of the
values of $\psi$ at the nodes (A,D,E,F), the value of $\Delta$ and
$\Xi$ at E, and $h$.  Once this linear system is
solved\footnote{Although this is simple to do, the explicit solution
is rather long and we omit it from the current discussion.} we know
the values of all the fields at H (accurate to order $h^3$).
Finally, we can use
\begin{equation}
    \psi_\text{G} = \tfrac{3}{2} \psi_\text{E} - \psi_\text{D} +
    \tfrac{1}{4} \psi_\text{H} + \tfrac{1}{4} \psi_\text{A} +
    \mathcal{O}(h^4),
\end{equation}
to get the value of $\psi$ at G. Note that the values of $\Delta$
and $\Xi$ at the nodes D and G have not entered the discussion; it
turns out that it is not necessary to keep track of the brane
degrees of freedom at non-vertex nodes.

\subsection{Evolution in the bulk}\label{sec:diamond evolution}

In addition to evolving $\psi$, $\Delta$ and $\Xi$ near the brane,
we also need to evolve $\psi$ in the bulk.  We can take the diamond
CIJF to be a typical bulk cell.  By simply integrating the bulk wave
equation (\ref{eq:psi EOM}) over this cell and using the divergence
theorem, we obtain
\begin{equation}
    2(\psi_\text{F} + \psi_\text{I} -
    \psi_\text{J} - \psi_\text{C}) = \int_\Diamond d^2 x \, V\psi.
\end{equation}
Using a bilinear approximation for the integrand yields
\begin{multline}\label{eq:diamond evolution}
    \psi_\text{J} = \psi_\text{F} + \psi_\text{I} - \psi_\text{C} +
    \tfrac{1}{16} \delta \u \, \delta \v [ V_\text{F} \psi_\text{F} +
    \\ V_\text{C} \psi_\text{C} + V_\text{I} \psi_\text{I} + V_\text{J}
    (\psi_\text{F} + \psi_\text{I} - \psi_\text{C}) ] + \mathcal{O}(h^4).
\end{multline}
Here, $\delta\u = \mathcal{O}(h)$ and $\delta\v = \mathcal{O}(h)$
are the dimensions of the cell in null coordinates.  Hence, given
the knowledge of $\psi$ on the past nodes F, I and C, we can obtain
the value at the node J accurate to order $h^3$.

\subsection{Initial data and computational algorithm}

Having obtained the formulae that tell us how to evolve the fields
across individual cells, we are now in a position to discuss our
overall computational strategy.  For simplicity, we will describe
how the calculation is carried out on the sparse grid shown in
Fig.~\ref{fig:boundary cells}, but the method is easily generalized
to the denser grids shown in Fig.~\ref{fig:grids}.

For the numeric solution of a conventional hyperbolic problem, it
would be sufficient to specify initial data for $\psi$ on the nodes
(A,B,C,I) in addition to $\Delta_\text{A}$ and $\Xi_\text{A}$.
However, due to the nonlocal boundary condition associated with DGP
perturbations, we need to also specify $\psi_\text{D}$,
$\psi_\text{E}$, $\Delta_\text{E}$, and $\Xi_\text{E}$ initially
(see \cite{Deffayet:2004xg} for a detailed discussion of
wellposedness and initial conditions for DGP perturbations). Once
the initial data has been selected, the algorithm proceeds as
follows:
\begin{enumerate}
    \item the diamond evolution formulae (\ref{eq:diamond
    evolution}) is then used to obtain $\psi$ at the nodes F and J;
    \item the triangle evolution algorithm developed in \S\ref{sec:triangle
    evolution} gives $\psi_\text{G}$, $\psi_\text{H}$,
    $\Delta_\text{H}$, and $\Xi_\text{H}$;
    \item Eq.~(\ref{eq:diamond evolution}) is then used to find
    $\psi_\text{K}$; and finally,
    \item the triangle algorithm gives the field values at the
    remaining nodes L and M.
\end{enumerate}
Obviously, if we have a larger grid than the one shown in
Fig.~\ref{fig:boundary cells}, steps 2 and 3 need to be iterated a
number of times.

For generic grids, the number of diamond cells in the grid will
scale as $1/h^2$ while the number of triangle cells goes like $1/h$.
Since the errors involved in the diamond and triangle evolution
formula are $\mathcal{O}(h^4)$ and $\mathcal{O}(h^3)$, respectively,
we obtain a final answer that is quadratically $\mathcal{O}(h^2)$
convergent.

\section{Tensor perturbations about a de Sitter brane}\label{sec:tensor results}

It is possible to analytically solve the tensor mode perturbation
equations given in \S\ref{sec:tensor mode eqns} for the case of pure
tension branes.  The governing bulk wave equation is separable in
Gaussian-normal coordinates, which allows for a Kaluza-Klein mode
decomposition of the gravitational wave amplitude $E$.  Within the
spectrum, one can always find a single discrete mode that is
normalizable in the bulk.  The 4-dimensional mass of this excitation
is given by \cite{Koyama:2005br, Koyama:2005tx}
\begin{equation}
    \frac{m^2_0}{H^2} =
    \begin{cases}
        \frac{3Hr_\c - 1}{(Hr_\c)^{2}}, & \epsilon = +1 \text{ and } Hr_\c > 2/3, \\
        0, & \epsilon = -1.
    \end{cases}
\end{equation}
In practical terms, the existence of this bound state means that for
any choice of initial data, we expect the late time behaviour of the
brane amplitude to be well described by a solution of
\begin{equation}
    \ddot E_\b + 3H \dot E_\b + \left( \frac{k^2}{a^2} + m_0^2
    \right) E_\b = 0,
\end{equation}
where $a = a(t) = e^{Ht}$. It is easy to verify that this implies
\begin{equation}
    \lim_{t\rightarrow\infty} E_\b(t) \propto
    \begin{cases}
        a^{-1/Hr_\c}, & \epsilon = +1 \text{ and } Hr_\c > 2/3, \\
        a^0, & \epsilon = -1.
    \end{cases}
\end{equation}
\begin{figure*}
\includegraphics[width=0.95\textwidth]{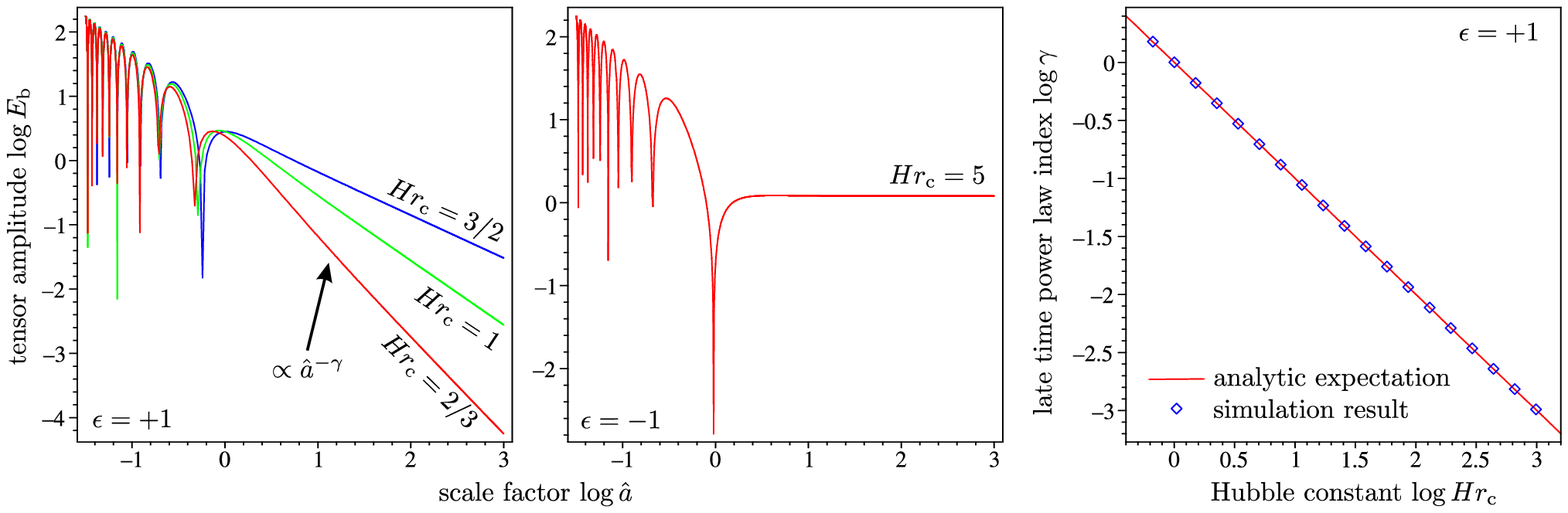}
\caption{Numeric solutions for the amplitude of tensor mode
perturbations $E_\b$ on a pure tension brane (left and center) and
comparison of the late time power law index derived from analytic
and simulation results (right). Note that our late time simulation
results for $\epsilon = -1$ are all very similar to the $Hr_\c = 5$
case shown here, and are all consistent with the analytic $\gamma =
0$ expectation.}\label{fig:tensor simulations}
\end{figure*}
In Fig.~\ref{fig:tensor simulations}, we plot a few typical results
for the on brane profile of $E$.  All simulations share the property
that they approach scaling solutions at late time:
\begin{equation}
    E_\b \approaches{t}{\infty} a^{-\gamma}.
\end{equation}
In all of our simulations for $\epsilon = -1$, we find that $\gamma
= 0$, which is consistent with the analytic expectation that the
normal branch spectrum contains a massless bound state.  On the
other hand, for the self-accelerating brane we would expect
\begin{equation}
    \gamma = \frac{1}{Hr_\c}, \quad \epsilon = +1.
\end{equation}
Also in Fig.~\ref{fig:tensor simulations}, we plot this
theoretical expectation versus our numerical results, and we see
excellent agreement. Hence, we have confirmed that our code
reproduces analytic results for the gravitational waves in the
pure tension DGP model.

\section{Scalar perturbations in the self-accelerating
universe}\label{sec:scalar SA}

\subsection{Cosmological parameters for DGP late-time
acceleration}

In this section, we concentrate on the $w = \sigma = 0$ and
$\epsilon = +1$ DGP scenario as a model for the late-time
accelerating universe.  Examining Table \ref{tab:coefficients} and
Eqns.~(\ref{eq:H and k def},\ref{eq:brane EOM},\ref{eq:scalar
potential}) in detail, we see that the entire evolution of scalar
perturbations (modulo initial data) in the canonical formalism is
governed by a single parameter $\rhohat_*$. Recall that we defined
$\rhohat_*$ to be proportional to the matter density when the mode
being modeled crosses the Hubble horizon. However, we note that in
any model which asymptotes to de Sitter space in the future, any
mode that enters the horizon must exit it as well. Indeed, by
solving the equation $\k/\H\a = 1$, we find that there are always
two horizon crossing epochs:
\begin{equation}
    \a = 1 \text{ and } \frac{2\rhohat_*}{3 + \sqrt{9 +
    12\rhohat_*}}.
\end{equation}
If $\rhohat_* > 6$, the second solution is always greater than the
first; i.e., the perturbation enters the horizon at $\a = 1$ and
leaves later.  However, if $\rhohat_* = 6$ both solutions coincide
and horizon entry and exit occur at the same time. It is easy to
confirm that the physical wavelength of the $\rhohat_* = 6$ mode
coincides with the Hubble length at the moment when the brane
switches from the decelerating to the accelerating phase; i.e., when
$\ddot a = 0$.

We can go further by incorporating data from cosmological
observations.  By examining probes of the expansion history
\cite{Lazkoz:2007zk}, it has been found that
\begin{equation}\label{eq:SA observational constraint}
    \Omega_{r_\c} = \frac{1}{4H_0^2 r_\c^2} = 0.15 \pm 0.02,
\end{equation}
at 95\% confidence.\footnote{Actually, in \cite{Lazkoz:2007zk} the
constraint $\Omega_m = 0.23 \pm 0.04$ was given, from which
(\ref{eq:SA observational constraint}) can be easily derived.} Once
the value of $\Omega_{r_\c}$ is fixed, we can use it to find the
value of $a_*$ for any given mode in terms of $\rhohat_*$:
\begin{equation}
    a_*^3 = \frac{1}{\a_0^3} = \frac{3(1-2\Omega_{r_\c}^{1/2})}{4\Omega_{r_\c}
    \rhohat_*}.
\end{equation}
This in turn allows us to determine the comoving wavenumber from
\begin{equation}\label{eq:k equation}
    k = \frac{\k a_*}{r_\c} = 2 \Omega^{1/2}_{r_\c} \k a_* H_0, \quad
    H_0^{-1} = 2998 \, h^{-1} \, \text{Mpc}.
\end{equation}
Hence, we can find $k$ in terms of $\rhohat_*$.  In order to
facilitate the comparison of our results with observations and the
literature, it is more convenient to invert the procedure to obtain
$\rhohat_*$ in terms of $k$.  However, this only works if $k$ is
above a critical value
\begin{equation}
    k_\c = 2 \Omega^{1/6}_{r_\c} (1 - 2\Omega_{r_\c}^{1/2})^{1/3}
    H_0 = 0.0003 \, h \, \text{Mpc}^{-1}.
\end{equation}
The last equality follows from taking the best fit value for
$\Omega_{r_\c}$, which is what we do from now on.  Any modes with $k
< k_\c$ have physical wavelengths larger than the Hubble length
throughout the cold dark matter and late time acceleration epochs,
and therefore are not modeled by our code.  Finally, it is sometimes
useful to have an explicit expression for the gravitational master
variable $\Omega_\b$ in terms of the canonical field $\psi_\b$. This
is:
\begin{equation}\label{eq:various Omega}
    \Omega_\b = \frac{a_*}{4H_0^2 \Omega_{r_\c}} \a^{3/2} \psi_\b =
    \frac{a_*}{4H_0^2 \Omega_{r_\c}} \hat\Omega_\b.
\end{equation}

\subsection{Typical waveforms}\label{sec:SA typical}

\begin{figure*}
\includegraphics{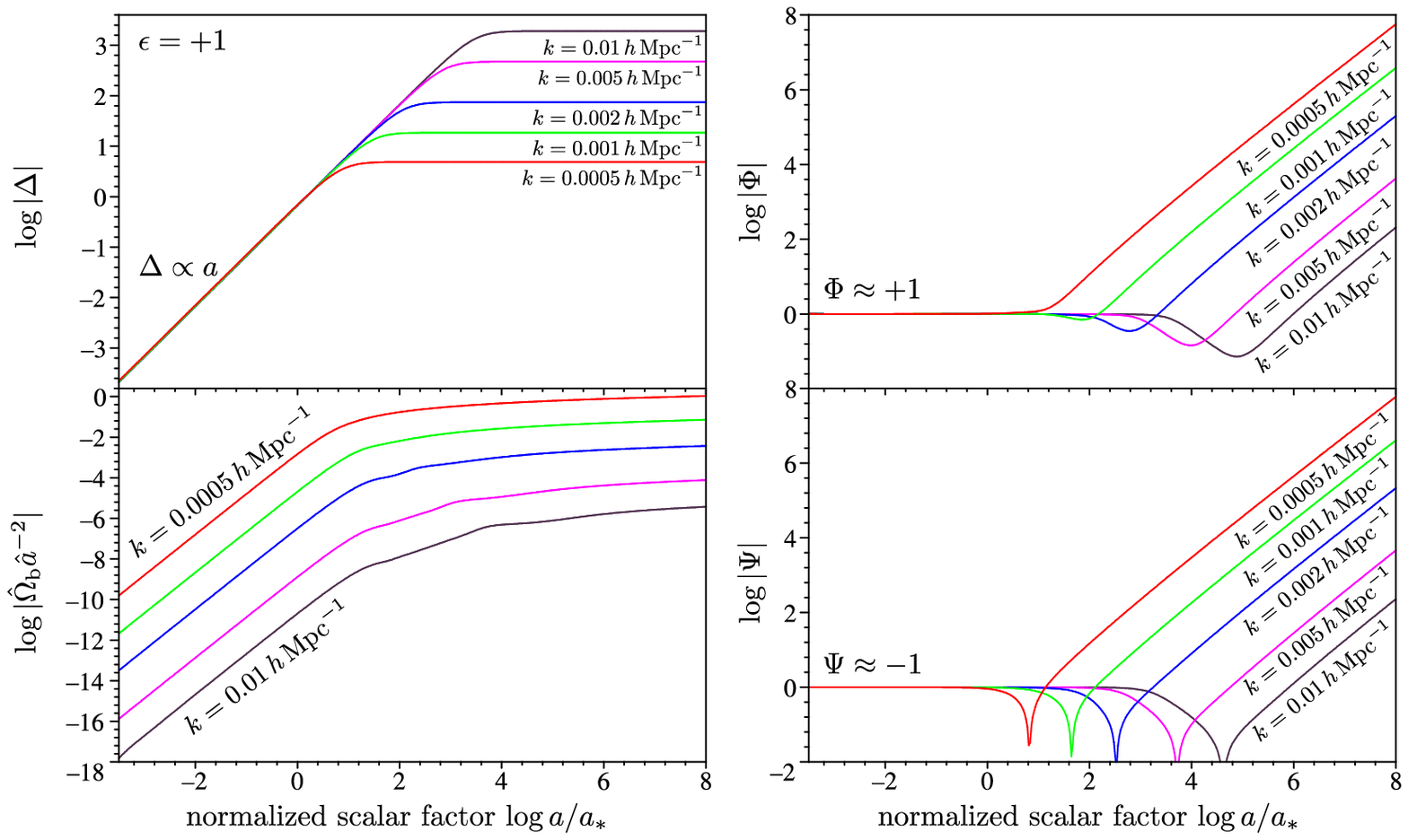}
\caption{The results of our simulations on the brane for several
choices of $k$.  We have normalized the value of $\Phi$ to be unity
at early times.  Also note that the lower left panel shows the
dimensionless bulk master variable $\hat\Omega_\b$, as defined in
Eq.~(\ref{eq:Omega hat def}), divided by $\a^2$.  All simulations
are performed with $\rhohat_* > 6$, which means that all modes
\emph{enter} the horizon at $\a = 1$, or when $a =
a_*$.}\label{fig:typical}
\end{figure*}

In Fig.~\ref{fig:typical}, we plot the results of our simulations
for several values of $k$ greater than the critical value $0.0003 \,
h \, \text{Mpc}^{-1}$.  For all plots of scalar perturbations in
this paper, we select the bulk field to be zero and the brane field
non-zero initially.  We have also simulated several different
choices of initial data, such as the bulk field being constant along
the initial null hypersurface, and have found that the simulation
results remain the same as long as the initial time is early enough.
This is analogous to what happens in the RS case
\cite{Cardoso:2007zh}.

From Fig.~\ref{fig:typical} it can be seen that we recover ordinary
4-dimensional GR at very early times. In particular, for $a \ll a_*$
and all values of $k > k_\c$ we see that:
\begin{itemize}
    \item the metric perturbations are conserved and have the
    opposite sign, $\Phi \approx -\Psi$;
    \item the density perturbation is proportional to the scale
    factor, $\Delta \propto a$; and,
    \item the bulk master variable scales as $\Omega_\b \propto
    a^4$ on the brane.
\end{itemize}
In addition, we have checked that the 4-dimensional Poisson
equation is satisfied before horizon entry:
\begin{equation}
    \frac{k^2}{a^2} \Phi - \frac{1}{2} \kappa_4^2 \rho \Delta
    \approx 0, \quad a \ll a_*.
\end{equation}
In other words, we have explicitly confirmed that DGP perturbations
behave as in GR on superhorizon scales before horizon crossing.

\begin{figure}
\includegraphics{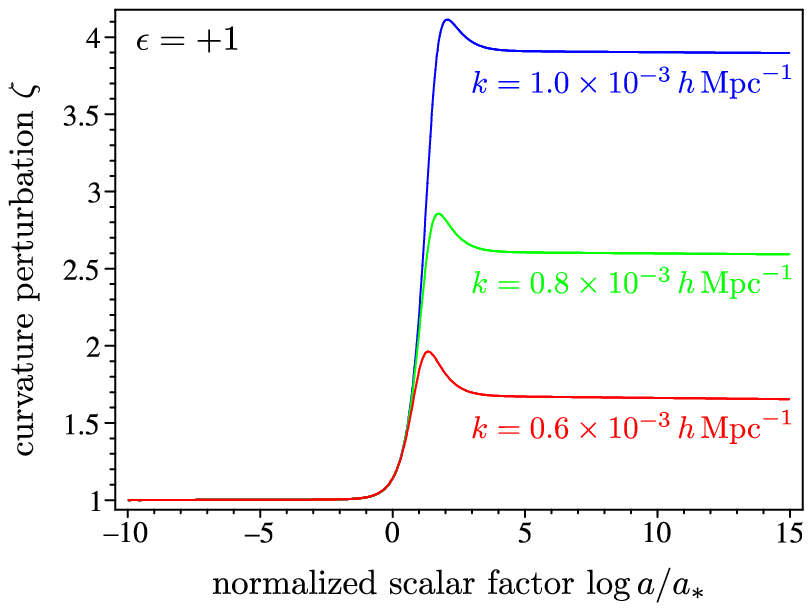}
\caption{Behaviour of the $\zeta$ curvature perturbation on large
scales for the self-accelerating branch (we have normalized $\zeta =
1$ at early times). Note that the curvature perturbation is
conserved when the modes are superhorizon; i.e., at both early and
late times. This is to be expected for any conservative theory of
gravity, such as the DGP model.}\label{fig:zeta}
\end{figure}

Finally, in Fig.~\ref{fig:zeta} we plot the simulation results for
the behaviour of the curvature perturbation $\zeta$ as given by
(\ref{eq:zeta}) for a few different large scales.  As can clearly be
seen, $\zeta$ is conserved for both early and late times when the
physical wavelengths of the modes are much larger than the horizon
size.  This is to be expected for any metric theory of gravity
\cite{Wands:2000dp}, and hence provides a good consistency check of
our code.

\subsection{The quasi-static
approximation and subhorizon behaviour}

In \cite{Koyama:2005kd}, a `quasi-static' (QS) approximation was
developed to describe the behaviour of DGP perturbations whilst
well inside the cosmological horizon and with physical wavelengths
much less than the crossover scale:
\begin{equation}
    k \gg Ha, \quad a \ll kr_\c.
\end{equation}
These conditions will hold for modes with $k \gg k_\c$ and $k \gg
2\Omega^{1/2}_{r_\c} H_0$ (up to some redshift); or, equivalently,
if
\begin{equation}
    k \gg 10^{-4} \, h \, \text{Mpc}^{-1}.
\end{equation}
In this section, we compare the QS approximation to our
simulations to determine just how large $k$ must be for it to be
valid.

In the QS approximation, one neglects the time derivatives of
$\Omega$ compared to the spatial gradients.  This allows one to
solve the bulk wave equation (\ref{eq:scalar master eqn}), and hence
close the system (\ref{eq:scalar boundary condition}) and
(\ref{eq:scalar brane equation}) on the brane.  This leads to the
following ordinary differential equation for $\Delta$:
\begin{equation}
    \ddot\Delta + 2H\Delta = \frac{1}{2} \kappa_4^2 \left( 1 +
    \frac{1}{3\beta} \right),
\end{equation}
where
\begin{equation}
    \beta = 1 - 2\epsilon Hr_\c \left( 1 + \frac{\dot H}{3H^2} \right).
\end{equation}
In addition, the following relations are predicted to hold:
\begin{subequations}\label{eq:QS Phi and Psi}
\begin{align}
    \Phi & = + \frac{\kappa_4^2 \rho a^2}{2k^2} \left( 1 -
    \frac{1}{3\beta} \right) \Delta , \\ \Psi & = - \frac{\kappa_4^2 \rho a^2}{2k^2} \left( 1
    + \frac{1}{3\beta} \right) \Delta.
\end{align}
\end{subequations}

\begin{figure}
\includegraphics{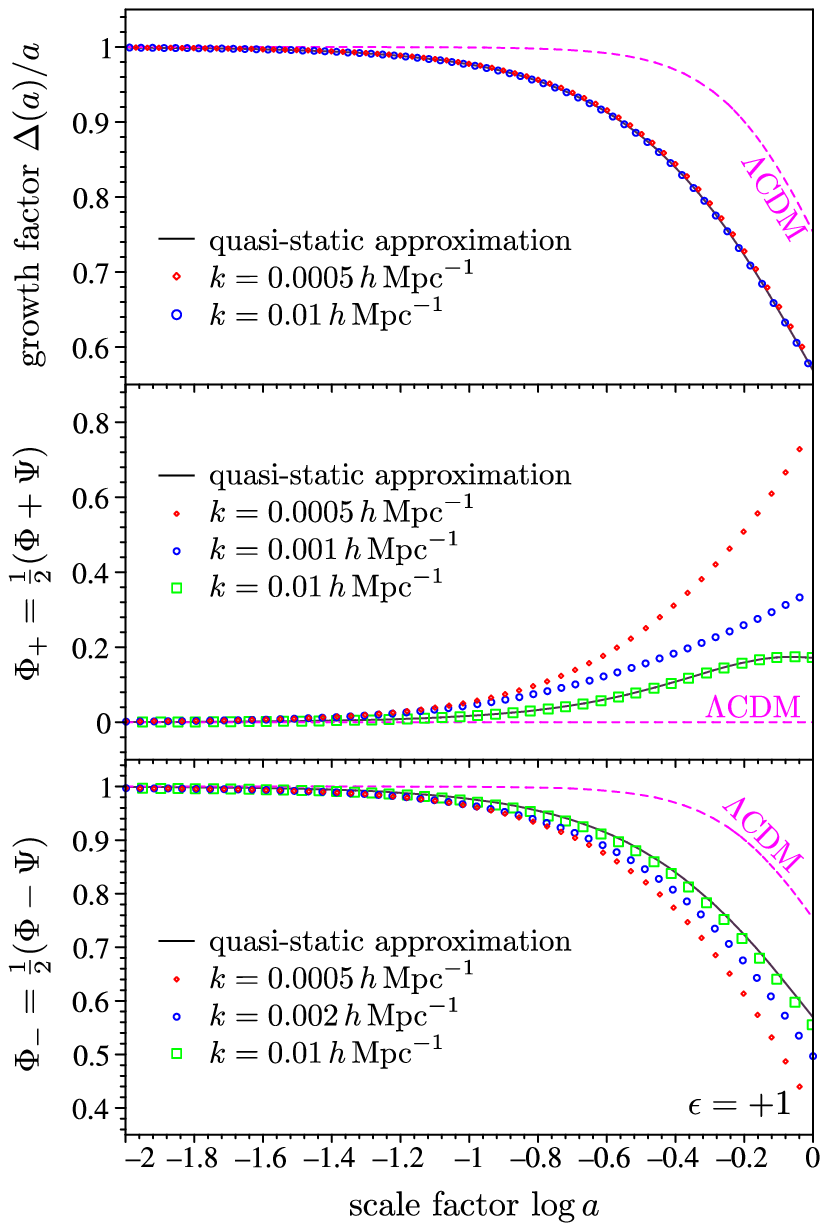}
\caption{Linear growth factor and alternate gravitational potentials
$\Phi_\pm$ from simulations and the QS approximation in the
self-accelerating branch. In the top panel, we normalize $g(a)$ to
unity at early times, in the lower two panels we normalize $\Phi_-$
to unity as $a \rightarrow 0$. For comparison, we also show the
relevant results for the concordance $\Lambda$CDM model with
${\Omega_m} = 0.26$ and $\Omega_\Lambda =
0.74$.}\label{fig:quasistatic compare 2}
\end{figure}

In Fig.~\ref{fig:quasistatic compare 2}, we compare simulation
results versus the QS approximation for the linear growth factor
$g(a) = \Delta(a)/a$, and the alternate gravitational potentials
$\Phi_\pm = \tfrac{1}{2}(\Phi \pm \Psi)$. We see that the simulation
results are consistent with the QS approximation for $k \gtrsim
10^{-2}\,h\,\text{Mpc}^{-1}$.

\begin{figure}
\includegraphics{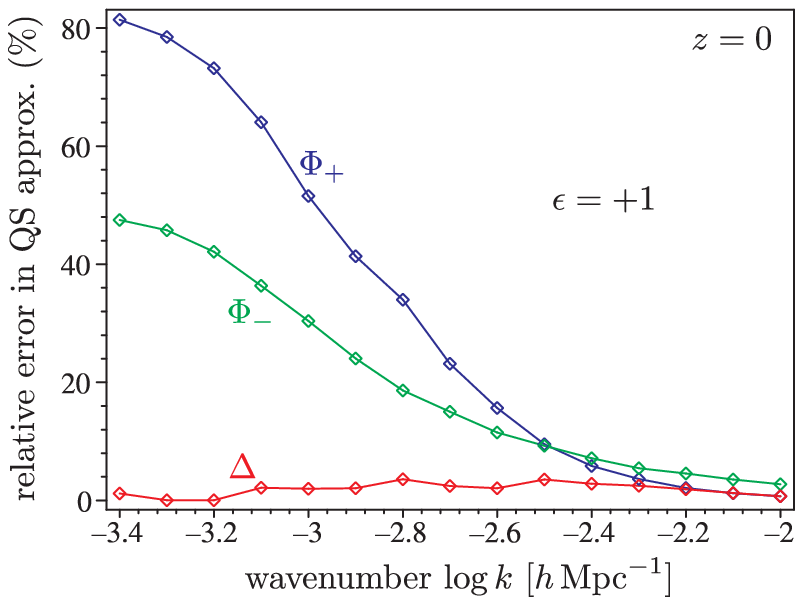}
\caption{The relative error in the QS approximation, as defined in
Eq.~(\ref{eq:relative error}), for various quantities evaluated in
the present epoch and assuming the self-accelerating branch.  For $k
\gtrsim 0.01 \, h \, \text{Mpc}^{-1}$ the errors are less than
5\%.}\label{fig:QS error}
\end{figure}

We further quantify the performance of the QS approximation in
Fig.~\ref{fig:QS error}.  There, we show the relative error in the
QS approximation as a function of the scale.  This is defined by
\begin{equation}\label{eq:relative error}
    \text{rel.~error} = \left| \frac{\text{QS prediction} -
    \text{simulation result}}{\text{simulation result}} \right|
    \times 100 \% .
\end{equation}
We see that the relative error in the QS prediction for $\Delta$
is fairly low ($<4\%$) on all scales.  Conversely, the QS values
of $\Phi_\pm$ become reliable only for $k \gtrsim
0.01 \,h\,\text{Mpc}^{-1}$, with errors of less than $\sim 5\%$.

\subsection{Superhorizon behaviour in the asymptotic
future}\label{sec:SA future}

In this subsection, we attempt to explain/predict the very late time
behaviour of our simulations by demonstrating that there exists a
bound state of the bulk field in the asymptotic future of the
evolution.  As we have already seen for the case of tensor
perturbations in \S\ref{sec:tensor results}, such bound states tend
to dominate the late time behaviour of the model, irrespective of
initial data.

In the asymptotic future, the brane geometry approaches that of de
Sitter space with $H = 1/r_\c$.  In the de Sitter regime, the bulk
wave equation (\ref{eq:scalar master eqn}) becomes
\begin{multline}\label{eq:asymptotic wave equation}
    0 = -\frac{\di^2\Omega}{\di t^2} + \frac{3}{r_\c}
    \frac{\di\Omega}{\di t} + \left( 1+\frac{y}{r_\c} \right)^2
    \frac{\di^2\Omega}{\di y^2} \\ -
    \frac{2}{r_\c} \left( 1+\frac{y}{r_\c} \right)  \frac{\di\Omega}{\di
    y} - \frac{k^2}{a^2} \Omega,
\end{multline}
with $a = e^{t/r_\c}$.  This equation is solvable via the separation
of variables $\Omega(t,y) = T_\lambda(t)\omega_\lambda(y)$, where
\begin{subequations}
\begin{align}
    0 & = \frac{d^2 T_\lambda }{dt^2} - \frac{3}{r_\c} \frac{dT_\lambda }{dt} + \left(
    \frac{k^2}{a^2} - \frac{\lambda}{r_\c^2} \right) T_\lambda, \label{eq:time equation} \\
    0 & = \frac{d}{dy} \left[ \frac{1}{(y+r_\c)^2} \frac{d\omega_\lambda}{dy} \right]
    - \frac{\lambda}{(y+r_\c)^4} \omega_\lambda.
\end{align}
\end{subequations}
Here, $\lambda$ is a dimensionless separation constant.  The
solution for $\omega_\lambda$ is
\begin{equation}
    \omega_\lambda = a_+ \left( 1 + \frac{y}{r_\c} \right)^{\nu_+} +
    a_- \left( 1 + \frac{y}{r_\c} \right)^{\nu_-},
\end{equation}
where $\nu_\pm = \frac{3}{2}\left(1 \pm \sqrt{1 + \frac{4}{9}
\lambda}\right)$. Assuming that $\lambda$ is real, we need to set
$a_+ = 0$ to ensure that $\omega_\lambda$ is normalisable under the
Sturm-Louisville inner product; i.e., that
\begin{equation}
    (\omega_\lambda, \omega_\lambda) = \int_0^\infty dy
    \frac{\omega_\lambda^2(y)}{(y+r_\c)^4},
\end{equation}
is finite, which means we have a true bound state.

Now, if we put $H = 1/r_\c$ in the boundary condition
(\ref{eq:scalar boundary condition}) we obtain
\begin{equation}\label{eq:asymptotic BC}
    (\di_y \Omega)_\b = -r_\c \left[ \ddot\Omega_\b - \frac{3}{r_\c} \dot\Omega_\b
    + \left( \frac{k^2}{a^2} + \frac{1}{r_\c^2} \right) \Omega_\b - \frac{2\kappa_4^2\rho
    a^3}{k^2} \Delta \right].
\end{equation}
Let us now assume that for very late times ($kr_\c \ll a$) the
following conditions hold:
\begin{equation}\label{eq:assumptions}
    \left| \frac{\kappa_4^2\rho
    a^3}{k^2} \Delta \right| \ll \min\left( |\ddot\Omega_\b|, r_\c^{-1} |\dot\Omega_\b|,
    r_\c^{-2} |\Omega_\b| \right);
\end{equation}
i.e., the $\Delta$ term is negligible on the righthand side of
(\ref{eq:asymptotic BC}).  Under these assumptions, which we still
need to justify, (\ref{eq:asymptotic wave equation}) and
(\ref{eq:asymptotic BC}) form a closed system for $\Omega$.  Putting
our mode solution $\Omega(t,y) = T_\lambda(t)\omega_\lambda(y)$ into
the boundary condition with $c_+ = 0$ and neglecting $\Delta$, we
obtain
\begin{equation}
    \lambda = -2.
\end{equation}
Putting $\lambda = -2$ into the temporal equation (\ref{eq:time
equation}) and solving for $T_\lambda$, we find that
\begin{equation}
    T_\lambda(t) \approaches{t}{\infty} b_1 a^2(t) + b_2 a(t),
\end{equation}
where $b_1$ and $b_2$ are constants.  Of course, the $b_1$ solution
will eventually dominate, which leads to the following asymptotic
bound state:
\begin{equation}\label{eq:Omega soln}
    \Omega(t,y) \approaches{t}{\infty} \Omega_0 \left( 1 + \frac{y}{r_\c} \right) a^2(t),
\end{equation}
where $\Omega_0$ is a constant. However, before we assume that the
late time behaviour of the system is indeed described by this
bound state, we must verify that the assumptions
(\ref{eq:assumptions}) under which it was derived are valid.  To
do so, we note that when $H = 1/r_\c$, the $\Delta$ equation of
motion (\ref{eq:scalar brane equation}) reduces to
\begin{equation}
    \ddot\Delta + \frac{2}{r_\c} \dot\Delta - \kappa_4^2 \rho \Delta = -\frac{k^4}{3a^5}
    \Omega_\b.
\end{equation}
Making use of (\ref{eq:Omega soln}), this equation can be solved
exactly. However, the full solution is complicated and not really
relevant, so we just quote the late time behaviour
\begin{equation}\label{eq:Delta soln}
    \Delta(t) \approaches{t}{\infty} \Delta_0,
\end{equation}
where $\Delta_0$ is a constant. With the solutions (\ref{eq:Omega
soln}) and (\ref{eq:Delta soln}), we see that the assumptions
(\ref{eq:assumptions}) are indeed satisfied at sufficiently late
time. Hence we have succeeded in finding an asymptotic bound state
that is expected to dominate the system's behaviour at late time.
Finally, note that we can use these asymptotic solutions for
$\Omega$ and $\Delta$ with (\ref{eq:Phi and Psi formulae}) to
obtain
\begin{equation}\label{eq:Phi and Psi solns}
    \Phi \approaches{t}{\infty} \frac{\Omega_0 a(t)}{2r_\c^2},
    \quad \Psi \approaches{t}{\infty} \frac{\Omega_0 a(t)}{2r_\c^2};
\end{equation}
i.e., $\Phi \approx \Psi$ at late time. We have verified that the
asymptotic solutions (\ref{eq:Omega soln},\ref{eq:Delta
soln},\ref{eq:Phi and Psi solns}) are realized in our simulations at
late times.

Before moving on, we would like to remark on the apparent
instability of the self-accelerating DGP model as indicated by the
divergence of $\Omega$, $\Phi$ and $\Psi$ in the asymptotic future.
This unstable mode corresponds to the radion, which is a physical
degree of freedom in the self-accelerating branch despite the fact
that we have only one brane \cite{Koyama:2005tx}.  It is well known
that in this case the radion has a negative mass squared $m^2 =
-4H^2$ and thus it is unstable \cite{Gen:2000nu}. However, as was
shown in Ref.~\cite{Gen:2002rb}, this is not a true gravitational
instability on the brane as it is possible to find a gauge in which
all metric perturbations remain finite.
%

\section{Scalar perturbations in the DGP normal
branch}\label{sec:scalar NB}

\subsection{Cosmological parameters for $\Lambda$DGP}
We now turn our attention to the behaviour of density perturbations
in the normal branch of the DGP model.  Unlike the $\epsilon = +1$
case, this branch does not naturally have a late time accelerating
phase.  So in order to be made consistent with observations, we must
allow for the brane to have a nonzero tension that acts as an
effective 4-dimensional cosmological constant (we call this the
$\Lambda$DGP model). Assuming that the matter sector is
CDM-dominated, the Friedmann equation for this scenario follows from
the general form (\ref{eq:background dynamics}) with $\epsilon = -1$
and $w=0$. The background dynamics has been compared to observations
of $H(z)$ in \cite{Lazkoz:2007zk}, who finds the following parameter
values:
\begin{equation}\label{eq:normal branch observational constraint}
    \Omega_\text{m} = \frac{\kappa_4^2 \rho_0}{3H_0^2} = 0.23\pm 0.04, \quad
    \Omega_{r_\c} = \frac{1}{4H_0^2 r_\c^2} \le 0.05,
\end{equation}
at $95\%$ confidence.  Here, $\rho_0$ is the present day CDM
density.  Note that the observationally preferred value of
$\Omega_{r_\c}$ is zero.  Since the DGP model goes over to GR in
this limit, this implies that $\Lambda$CDM gives a better fit to the
data than $\Lambda$DGP.  In what follows, we will always assume the
best fit value of 0.23 for $\Omega_\text{m}$ and treat
$\Omega_{r_\c}$ as an adjustable parameter that must be smaller than
0.05 to yield a realistic model.

Once $\Omega_\text{m}$ and $\Omega_{r_\c}$ have been selected, it is
straightforward to obtain the value of the dimensionless brane
tension:
\begin{equation}
    \hat\sigma = \kappa_4^2 r_\c^2 \sigma =
    \frac{3( 1 - \Omega_\text{m} + 2\Omega_{r_\c}^{1/2}
    )}{4\Omega_{r_\c}},
\end{equation}
which can be re-written in terms of a new density parameter
$\Omega_\sigma$:
\begin{equation}
    \Omega_\sigma = \frac{\kappa_4^2\sigma}{3H_0^2} = 1 -
    \Omega_\text{m} + 2\Omega_{r_\c}^{1/2}.
\end{equation}
One can also find $a_*$ in terms of the observational parameters and
$\rhohat_*$:
\begin{equation}
    a_*^3 = \frac{1}{\a_0^3} =
    \frac{3\Omega_\text{m}}{4\Omega_{r_\c}\rhohat_*}.
\end{equation}
As before, these two formulae can be used in (\ref{eq:k equation})
to determine $k$ as an explicit function of $\rhohat_*$, or
$\rhohat_*$ as an implicit function of $k$.  Following the latter
approach means that when we select $k$ along with $\Omega_\text{m}$
and $\Omega_{r_\c}$, the evolution of perturbations is completely
specified up to the choice of initial data.

\subsection{Simulation results and comparison to the QS approximation}

\begin{figure}
\includegraphics{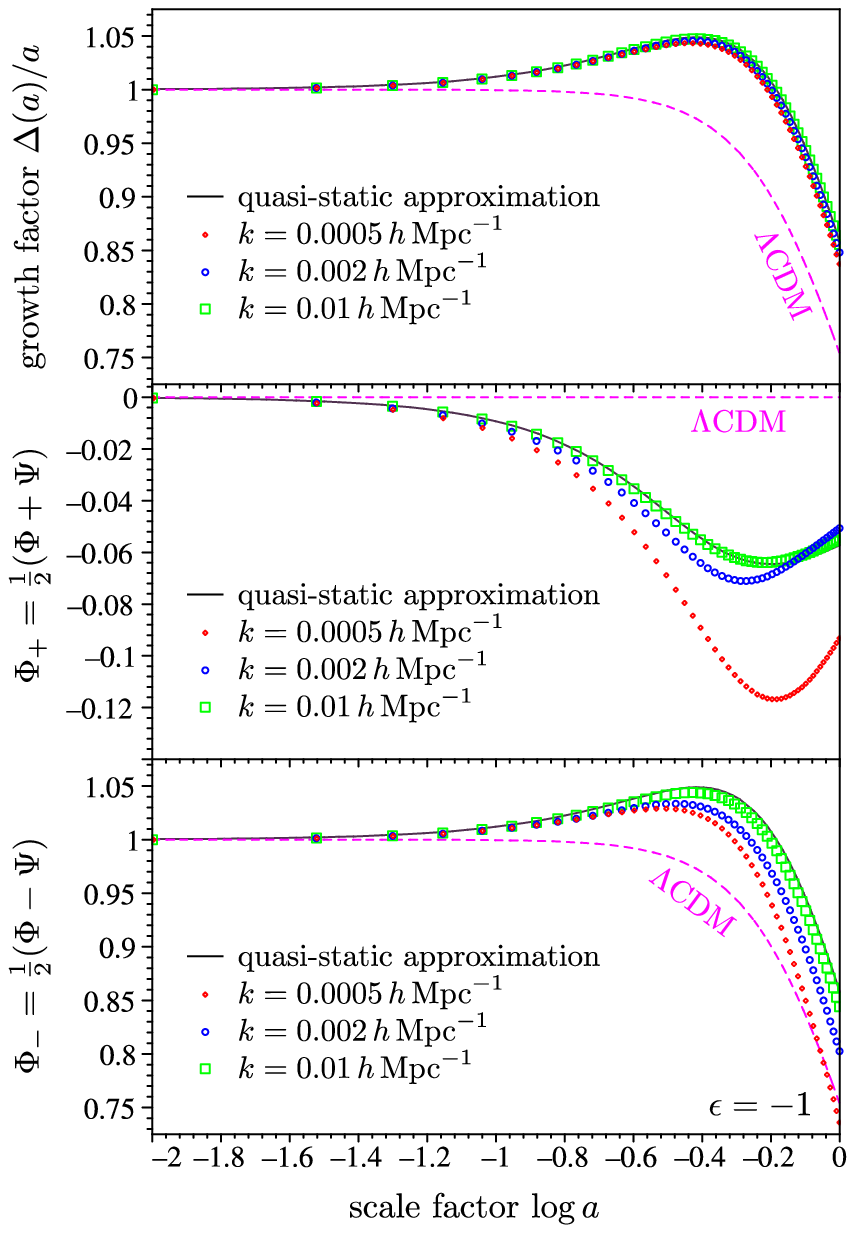}
\caption{Linear growth factor and alternate gravitational potentials
$\Phi_\pm$ from simulations and the QS approximation in the normal
branch with $\Omega_{r_\c} = 0.05$ and $\Omega_\text{m} = 0.23$. As
in Fig.~\ref{fig:quasistatic compare 2}, we normalize $g(a)$ and
$\Phi_-$ to unity at early times. For comparison, we also show the
relevant results for the concordance $\Lambda$CDM model with
$\Omega_\text{m} = 0.26$ and $\Omega_\Lambda =
0.74$.}\label{fig:quasistatic compare NB}
\end{figure}

In Fig.~\ref{fig:quasistatic compare NB}, we compare the results of
our simulations to the QS approximation and $\Lambda$CDM in the case
$\Omega_{r_\c} = 0.05$.  As in \S\ref{sec:SA typical}, we find that
the simulation results are fairly insensitive to initial conditions
provided that the initial data surface is set far enough into the
past; here, all plots have been generated assuming $\Omega = 0$
initially. In contrast to the self-accelerating case, we find that
the linear growth factor and $\Phi_-$ potential are generally larger
than in the $\Lambda$CDM case.  The general trend is for $\Phi_-$ to
become larger on small scales.  We also notice that the QS
approximation seems to provide a very good match to the simulation
results for $\Delta$ on all scales.

\begin{figure*}
\includegraphics{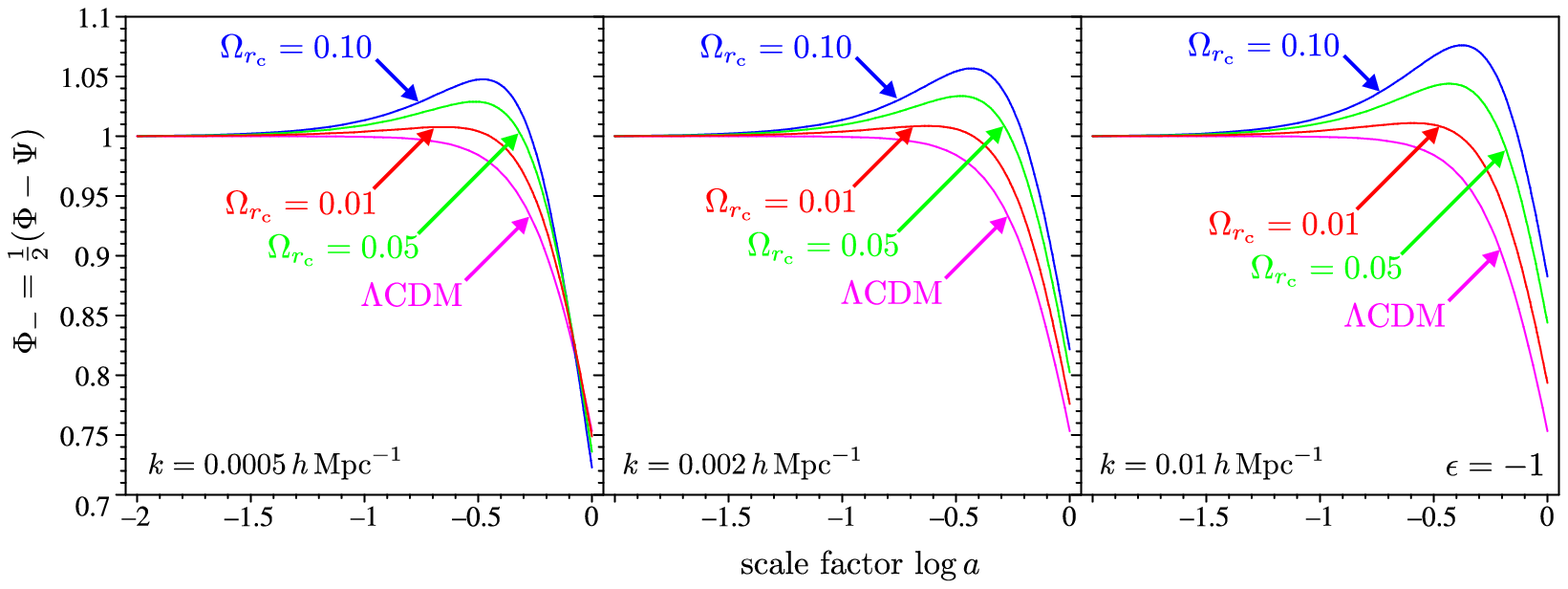}
\caption{The `ISW potential' $\Phi_-$ as a function of the scale
factor for various scales and choices of $\Omega_{r_\c}$ for the
normal branch.  In all cases we have taken $\Omega_\text{m}=0.23$.
The $\Lambda$CDM curves are included for purposes of
comparison.}\label{fig:quasistatic compare NB 2}
\end{figure*}

In Fig.~\ref{fig:quasistatic compare NB 2}, we show the effect of
changing the $\Omega_{r_\c}$ parameter on the simulation results for
$\Phi_-$.  For any given scale, we see that the $\Omega_{r_\c}
\rightarrow 0$ limit approaches the $\Lambda$CDM prediction.  Also,
we note that the simulation results are closer to the $\Lambda$CDM
case for smaller values of $k$; i.e., the most pronounced deviations
from GR are observed on the smallest scales simulated.

\begin{figure}
\includegraphics{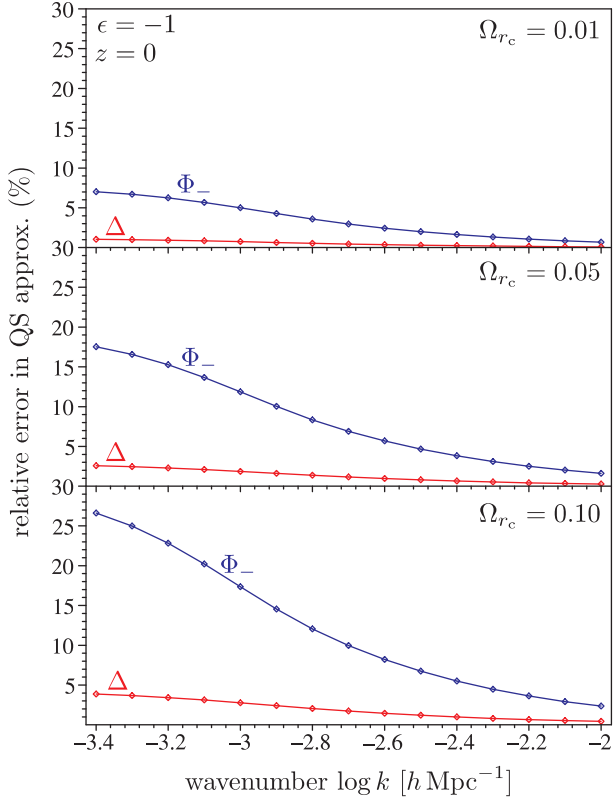}
\caption{The relative error in the QS approximation, as defined in
Eq.~(\ref{eq:relative error}), for various quantities evaluated in
the present epoch and assuming the normal branch.  For $k \gtrsim
0.01 \, h \, \text{Mpc}^{-1}$ the errors are less than 5\%. Also
notice how the QS approximation is generally more accurate for
smaller $\Omega_{r_\c}$.}\label{fig:QS error NB}
\end{figure}

Finally, in Fig.~\ref{fig:QS error NB} we quantify the error in the
QS prediction for the value of various quantities at $z = 0$ as a
function of the scale. As in the self-accelerating case, we see that
the QS approximation provides reasonably accurate results (with
errors $\lesssim 5\%$) on scales $k \gtrsim 0.01\, h\,
\text{Mpc}^{-1}$.

\subsection{Superhorizon behaviour in the asymptotic future}

In the asymptotic future, the brane geometry approaches that of de
Sitter spacetime with $H$ determined by $\sigma \ne 0$. Unlike in
the self-accelerating branch, there appears a horizon at $y=1/H$. An
analysis similar to the one presented in \S\ref{sec:SA future} for
the self-accelerating branch shows that there is no solution with a
real $\lambda$ \cite{Koyama:2006mh}.  Therefore, there is no bound
state solution in the asymptotic de Sitter spacetime and $\Omega$ is
a superposition of massive Kaluza-Klein modes that oscillate in
time.  It is worth noting that the dynamical scaling \emph{ansatz}
implicitly assumes the existence of a $\Omega$ bound state, and will
hence fail in the asymptotic de Sitter future of the DGP normal
branch.

\section{Conclusions}\label{sec:conclusions}

In this paper, we have presented numeric solutions for cosmological
perturbations in the DGP braneworld model both in the
self-accelerating and the normal branches.  We extended the
algorithm developed for the Randall-Sundrum (RS) model to handle the
nonlocal boundary conditions characteristic of the DGP model. The
numerical code was tested for tensor perturbations and the agreement
with the analytic solutions was found to be excellent.

We confirmed that on small scales $k > 0.01 h$ Mpc${}^{-1}$, the
quasi-static (QS) approximation reliably predicts the evolution of
perturbations with relative errors less than around $5\%$ at $z=0$.
Our results are quite insensitive to the initial conditions as long
as we start our simulations early enough.  On larger scales, the
potential $\Phi_-$, which determines the integrated Sach-Wolfe (ISW)
effect, shows more suppression than the QS prediction. We find that
our numerical solutions agree well with the dynamical scaling (DS)
solution both in the self-accelerating and normal branches, except
in the asymptotic de Sitter phase of the normal branch where the
dynamical scaling solution fails to exist \cite{Song}.

Our numeric solutions provide the basis for studying observational
signatures of the model, especially in the normal branch where the
influence of the extra dimension on the evolution of large scale
structure has not yet been explored.

\begin{acknowledgments}
We would like to thank Roy Maartens and Yong-Seon Song for many
useful conversations. AC is supported by FCT (Portugal) PhD
fellowship SFRH/BD/19853/2004. KK is supported by STFC (UK). SSS is
supported by STFC (UK) and NSERC (Canada). FPS is supported by FCT
(Portugal) PhD fellowship SFRH/BD/27249/2006.
\end{acknowledgments}

\appendix

\section{Boundary condition and $\Delta$ equation of motion for scalar
perturbations}\label{sec:scalar derivation}

To derive the boundary condition satisfied by the bulk master
variable $\Omega$ and the second order equation of motion for the
density perturbation $\Delta$ for scalar perturbations, we begin
with the linearized version of the effective Einstein equations
(\ref{eq:4D Einstein equations}):
\begin{equation}\label{eq:perturbed 4D Einstein equations}
    \delta G^{(4)}_{\mu \nu} =(2 \kappa_4^2 r_\c)^2 \delta\Pi_{\mu \nu}-{\delta \mathcal{E}}_{\mu
    \nu}.
\end{equation}
In this expression, $\delta G^{(4)}_{\mu \nu}$ and
$\delta\Pi_{\mu\nu}$ can be obtained by using the perturbed brane
metric (\ref{eq:perturbed induced metric}) and perturbed stress
energy tensor (\ref{eq:perturbed stress energy tensor}).  We can
parameterize the perturbations of the bulk Weyl ``fluid''
$\mathcal{E}_{\mu\nu}$ as follows:
\begin{equation}
\delta { \mathcal{E}}^{\mu}_{\,\,\nu} = -\kappa_4^2 \left(
\begin{array}{ccc}
-\delta \rho_{{ \mathcal{E}}} & & a \delta q_{{ \mathcal{E}},i} \\
a^{-1} \delta q_\mathcal{E}^{\,\,\,,i} &  & \tfrac{1}{3} \delta
\rho_{{ \mathcal{E}}} \: \delta^{i}_{\,\,j}
+ \delta \pi^{i}_{\mathcal{E} \,j}\\
\end{array}
\right).
\end{equation}
Here, $\delta\pi^\mathcal{E}_{ij}=\delta\pi^\mathcal{E}_{,ij}-
{1\over3}\delta\pi^{\mathcal{E},k}_{,k}\delta_{ij}$, a comma denotes
partial differentiation and indices are raised and lowered with the
flat 3-metric.  Using this in the $(0,i)$ component of the perturbed
effective Einstein equations (\ref{eq:perturbed 4D Einstein
equations}), we obtain the following equation:
\begin{equation}\label{eq:effective Einstein 1}
    H\Psi - \dot\Phi =
    \frac{\kappa_4^2}{2} \left( \frac{2 H r_\c \epsilon}
    {2 H r_\c \epsilon -1} \right) \left( \frac{\rho V}{k} +
    \frac{\delta q_\mathcal{E}}{2 H r_\c \epsilon} \right),
\end{equation}
where we have made use of the fact that the background brane matter
distribution is CDM plus a possible effective cosmological constant
induced by the brane tension.  Combining this with the $(0,0)$
component of (\ref{eq:perturbed 4D Einstein equations}) yields the
Poisson equation:
\begin{equation}\label{eq:poisson}
    \frac{k^2}{a^2} \Phi = \frac{\kappa_4^2}{2}
    \left( \frac{2\epsilon H r_\c}{2 \epsilon H r_\c -1} \right) \left( \rho\Delta
    - \frac{\delta\rho_\mathcal{E} - 3 H \delta q_\mathcal{E}}{2\epsilon H r_\c} \right).
\end{equation}
In these formulae, the gauge invariant density perturbation $\Delta$
is defined in (\ref{eq:Delta defn}), while the invariant velocity
perturbation is given by $V = -k\,\delta q/\rho$ in the longitudinal
gauge.

Additional relations can be obtained by noting that
\begin{equation}
    \delta(\nabla^\alpha T_{\alpha\beta}) = 0.
\end{equation}
The spatial and temporal components of this yield that
\begin{subequations}\label{eq:conservation equations}
\begin{gather}
    \dot V + H V =  \frac{k}{a}\Psi, \\
    \dot \Delta = -\frac{k}{a} \left( 1 - \frac{3a^2}{k^2}\dot H \right) V
    -3 ( \dot\Phi - H \Psi ),
\end{gather}
\end{subequations}
respectively.  Combining (\ref{eq:effective Einstein 1}) with
(\ref{eq:conservation equations}) yields a second order differential
equation for $\Delta$:
\begin{equation}
\label{eq:intermediate Delta eq}
    \ddot \Delta + 2 H \dot \Delta = -\frac{k^2}{a^2}\Psi + \frac{3}{2}\dot F + 3 H F,
\end{equation}
 with
\begin{equation}
     F = \frac{\kappa_4^2 \delta q_\mathcal{E}}{2 H r_\c \epsilon -1}.
\end{equation}

Now, one important feature that distinguishes braneworld
cosmological fluctuations from the GR case is that in addition to
perturbations of the geometry and the matter, we must also consider
perturbations of the brane's position.  That is, in the Gaussian
normal coordinates system the brane is located at $y = 0$ before
perturbation and at $y = \xi$ after perturbation, where $\xi$ is the
scalar brane bending degree of freedom.  It is useful to
parameterize the perturbed geometry of the $y = 0$ hypersurface
(i.e.,~the brane's unperturbed position) by
\begin{equation}\label{eq:perturbed induced metric y=0}
    ds_{y=0}^2 = -(1+2\mathcal{A})dt^2 + a^2 (1+2\mathcal{R})\delta_{ij} dx^i
    dx^j.
\end{equation}
The metric potentials at the unperturbed brane position
$(\mathcal{A},\mathcal{R})$ are then related to the metric
potentials at the perturbed position $(\Psi,\Phi)$ by
\begin{equation}\label{eq:metric potential transform}
    \Psi = \mathcal{A} - \epsilon \left( \frac{\dot H}{H} + H
    \right)\xi, \quad \Phi = \mathcal{R} - \epsilon H \xi.
\end{equation}
\citet{Deffayet:2002fn} has shown that the brane bending scalar is
simply given by
\begin{equation}
    \xi = -r_\c(\Phi + \Psi).
\end{equation}
In addition, he demonstrated that it is possible to express
$\mathcal{A}$ and $\mathcal{R}$ in terms of the bulk master variable
$\Omega$:
\begin{eqnarray}\nonumber
    \mathcal{A} &=& \frac{1}{6a} \left[ 3 \epsilon \left( \frac{\dot H}{H}
    - H \right)(\di_y \Omega)_\b + \frac{2k^2}{a^2}\Omega_\b - 3\ddot \Omega_\b +
    6 H \dot \Omega_\b \right],\\
    \mathcal{R} &=& \frac{1}{6a} \left[ 3\epsilon H (\di_y
    \Omega)_\b
    + \frac{k^2}{a^2}\Omega_\b - 3 H\dot\Omega_\b \right].
    \label{eq:A and R in terms of Omega}
\end{eqnarray}
It can also be shown \cite{Deffayet:2002fn} that the Weyl fluid
perturbations are also directly given by $\Omega$:
\begin{equation}\label{eq:Weyl fluid}
    \kappa_4^2 \delta \rho_\mathcal{E} = \frac{k^4 \Omega}{3 a^5}, \quad
    \kappa_4^2 \delta q_\mathcal{E} = -\frac{k^2}{3 a^3}(H\Omega-\dot\Omega).
\end{equation}
Now, these formulae in conjunction with the wave equation
(\ref{eq:scalar master eqn}) can be used to re-write the Poisson
equation (\ref{eq:poisson}) as:
\begin{equation}
    2r_\c\kappa_4^2 \rho \Delta = 4 r_\c \frac{k^2}{a^2}\Phi +
    \frac{2k^2}{a^2} \xi -\frac{k^2}{a^3}[(\di_y\Omega)_\b -\epsilon H
    \Omega_\b].
\end{equation}
Then, one can use (\ref{eq:metric potential transform})--(\ref{eq:A
and R in terms of Omega}) in this equation to obtain the boundary
condition (\ref{eq:scalar boundary condition}).  One can then use
(\ref{eq:scalar boundary condition}) with (\ref{eq:metric potential
transform})--(\ref{eq:A and R in terms of Omega}) to get $\Phi$ and
$\Psi$ in terms of $\Omega_\b$ and $\Delta$; i.e., Eqs.~(\ref{eq:Phi
and Psi formulae}). Finally, substituting (\ref{eq:Phi and Psi
formulae}) and (\ref{eq:Weyl fluid}) into (\ref{eq:intermediate
Delta eq}) gives the final form of the $\Delta$ equation of motion
(\ref{eq:scalar brane equation}).

\bibliography{DGP_perts}

\end{document}